\documentclass[preprint,showpacs,preprintnumbers,amsmath,amssymb,eqsecnum]{revtex4}

\newcommand{\sq}{\square}

\usepackage{graphicx}
\usepackage{dcolumn}
\usepackage{bm}
\usepackage[mathscr]{euscript}

\begin{document}

\setlength{\baselineskip}{6.65mm}


\title{Canonical Formalism for a $\boldsymbol{2n}$-Dimensional Model 
\\ with Topological Mass Generation}

\author{Shinichi Deguchi} 
\email{deguchi@phys.cst.nihon-u.ac.jp} 
\affiliation{%
Institute of Quantum Science, College of Science and Technology, 
Nihon University, Chiyoda-ku,  Tokyo 101-8308, Japan
}%


\date{\today}

\begin{abstract}
The four-dimensional model with topological mass generation that was found by 
Dvali, Jackiw and Pi has recently been generalized to any even number of 
dimensions ($2n$-dimensions) in a nontrivial manner 
in which a St\"{u}ckelberg-type mass term is introduced 
[S. Deguchi and S. Hayakawa, Phys. Rev. D {\bf 77}, 045003 (2008), 
arXiv:hep-th/0711.1446]. 
The present paper deals with a self-contained model, called here a modified hybrid model, 
proposed in this $2n$-dimensional generalization 
and considers the canonical formalism for this model. 
For the sake of convenience, 
the canonical formalism itself is studied for a model equivalent to 
the modified hybrid model by following the recipe for treating constrained 
Hamiltonian systems. 
This formalism is applied to the canonical quantization of the equivalent model 
in order to clarify observable and unobservable particles in the model. 
The equivalent model (with a gauge-fixing term) is converted to the modified hybrid model 
(with a corresponding gauge-fixing term) 
in a Becchi-Rouet-Stora-Tyutin (BRST)-invariant manner.  
Thereby it is shown that the Chern-Pontryagin density behaves as an observable massive particle 
(or field). The topological mass generation is thus verified at the quantum-theoretical level. 
\end{abstract}

\maketitle



\section{\label{sec:level1}Introduction} 

Various mass-generation mechanisms 
have been studied in classical and quantum field theories. 
Some of these mechanisms can be described in topological terms, 
in which topological entities play essential roles. 
For instance, 
in the topologically massive gauge theory in three dimensions \cite{DJT}, 
a Chern-Simons term included in the action 
makes gauge fields massive. 
In the four-dimensional analogue of this theory \cite{Lah, BCS, Deg},  
a topological entity called $BF$ term plays a role of the Chern-Simons term  
in generating masses of gauge fields. 
The topologically massive gauge theories thus describe  
mass-generation phenomena of vector fields.

A four-dimensional model with mass generation 
that is recently presented by Dvali, Jackiw and Pi  \cite{DJP}  
is also formulated in topological terms using topological entities:  
Chern-Pontryagin density $\mathcal{P}$ and Chern-Simons current 
$\mathcal{C}^{\mu}$, $\mathcal{P}=\partial_{\mu} \mathcal{C}^{\mu}$. 
Dvali {\it et al.} found their model as a partial, four-dimensional generalization 
of the (bosonized) Schwinger model \cite{Sch} reformulated in terms of 
$\mathcal{P}$ and $\mathcal{C}^{\mu}$ in two dimensions. 
Unlike the topologically massive gauge theories, the Dvali-Jackiw-Pi  (DJP) model 
describes mass generation of a pseudoscalar degree of freedom.  
In addition, the DJP model needs the presence of the chiral anomaly to generate  
a mass gap. Also, the action of the DJP model contains higher dimensional terms 
with respect to gauge fields. 
Therefore the DJP model is essentially different 
from the topologically massive gauge theories, although they share common topological 
terms.

Recently, the DJP model in four dimensions has been generalized to any 
even number of dimensions (or simply $2n$ dimensions) \cite{DH}. 
There, it was demonstrated that the topological mass generation studied by 
Dvali {\it et al.} is valid in $2n$ dimensions with no essential changes. 
As in the four-dimensional model, 
the presence of the chiral anomaly is crucial to  
this mass-generation mechanism.  
In Ref.~\onlinecite{DH}, another $2n$-dimensional model 
with topological mass generation was also proposed. 
In this model, a St\"{u}ckelberg-type mass term gives rise to mass generation 
of a pseudoscalar degree of freedom in a gauge invariant manner. 
In addition, a hybrid of the $2n$-dimensional models mentioned above 
was considered, in which generating a mass is caused by both the St\"{u}ckelberg-type  
mass term and the presence of the chiral anomaly. 
Because the hybrid model involves the St\"{u}ckelberg-type model and the DJP model 
as particular cases, it is sufficient to examine only the hybrid model.

The hybrid model, as well as the DJP model, is, however, not self-contained  
in the sense that the presence of the chiral anomaly is {\it a priori} assumed in the model  
without specifying its origin. For this reason, it is difficult 
to investigate definite properties of the hybrid model in its present form. 
By making some modification of the hybrid model, it becomes possible to derive    
the chiral anomaly within the framework of the hybrid model, without 
setting extra assumptions (see Sec. 5 of Ref.~\onlinecite{DH}).  
In this way, the hybrid model is promoted to a self-contained model.  
The modified model consists of a pseudoscalar field, $\eta$, 
and an antisymmetric pseudotensor field, $p^{\mu\nu}$, together with the 
topological entities $\mathcal{P}$ and $\mathcal{C}^{\mu}$ in $2n$ dimensions.  
The Lagrangian of this model is given in Eq. (\ref{1.7}) below.  
It is remarkable that the Yang-Mills fields constituting 
$\mathcal{P}$ and $\mathcal{C}^{\mu}$ 
appear in the equations of motion 
in the modified hybrid model only through $\mathcal{P}$ and $\mathcal{C}^{\mu}$.

In this paper, we investigate particle contents of the modified hybrid model, 
clarifying observable and unobservable particles. 
To this end, we consider the canonical formalism of 
a model equivalent to the modified hybrid model. 
The equivalent model is governed by a Lagrangian that has the same form as 
the Lagrangian of the modified hybrid model,  
but does not contain the constituent Yang-Mills fields 
(see Eq. (\ref{1.12}) below). 
The Chern-Simons current $\mathcal{C}^{\mu}$ 
in the modified hybrid model can be treated there as a fundamental field. 
For this reason, it is possible to make the investigation using the equivalent model.

The equivalent model possesses an Abelian gauge symmetry with a tensorial gauge 
parameter, and hence it is necessary to carry out gauge fixing for this symmetry  
to study the quantum-mechanical properties of the model. 
Although the gauge symmetry in question is Abelian,  
we adopt the gauge-fixing procedure based on 
the Becchi-Rouet-Stora-Tyutin (BRST) invariance principle 
(or simply BRST gauge-fixing procedure) \cite{KU, NO}. 
The BRST invariance principle is 
useful not only for determining gauge-fixing and Faddeev-Popov (FP) ghost terms 
but also for converting the equivalent model into the modified hybrid model  
in a BRST-invariant manner. In fact, the equivalent model becomes 
the modified hybrid model by adding a BRST-coboundary term to 
the Lagrangian of the equivalent model.

After carrying out the gauge fixing in the equivalent model, 
we consider the canonical formalism of this model 
by following the recipe for treating constrained Hamiltonian systems 
\cite{Dir, HRT, HT}.  
On detailed analysis of the constraints in phase space, 
it is shown that the equivalent model (with a gauge-fixing term),  
which originally contains antisymmetric pseudotensor fields,  
can be described only 
in terms of pseudoscalar fields supplemented with a modified Poisson bracket. 
The canonical quantization of the equivalent model is performed on  
the Hamiltonian system consisting only of the pseudoscalar fields. 
These fields are quantized with the canonical (anti-)commutation relations 
based on the modified Poisson bracket. 
In accordance with the BRST transformation rules of the pseudoscalar fields, 
each of the fields is classified into  
a BRST singlet or quartet field, and only the BRST-singlet fields are 
recognized to be genuinely physical in the sense of the Kugo-Ojima criterion 
\cite{KO, NO}. 
Noting the commutation relations concerning the BRST-singlet fields,  
we see that the only massive pseudoscalar field which can be observed with 
finite probability is present in the equivalent model. 
From the aspect of the modified hybrid model,  
the presence of a massive pseudoscalar field 
is understood as a topological mass-generation phenomenon. 
In fact, the Chern-Pontryagin density $\mathcal{P}$ is shown to behave as 
an observable massive field.

This paper is organized as follows. 
Section 2 introduces the topological entities and provides a brief review of 
the modified hybrid model. The equivalent model is also presented there. 
Section 3 treats the gauge and BRST symmetries of the equivalent model. 
The BRST gauge-fixing procedure is also considered after setting  
an appropriate gauge-fixing condition. 
Section 4 studies the canonical formalism of the equivalent model 
by following the recipe for treating constrained Hamiltonian systems. 
Section 5 performs the canonical quantization of the equivalent model 
by utilizing the results obtained in Sec. 4 and investigates particle contents 
of the model. 
Section 6 presents a BRST-invariant procedure for converting the equivalent model into 
the modified hybrid model and makes sure of the topological mass generation at the 
quantum-theoretical level. 
Section 7 is devoted to a summary and discussion.

\section{\label{sec:level1}A $\boldsymbol{2n}\,$-dimensional model 
with topological mass generation and its equivalent model} 

Let $A_{\mu}$ be a (Hermitian) Yang-Mills field on $2n$-dimensional 
Minkowski space, $\mathbf{M}^{2n}$, with Cartesian coordinates $(x^{\mu})$.  
The field $A_{\mu}$ is assumed to take values in a compact  
semisimple Lie algebra $\frak{g}$, and hence $A_{\mu}$ can be expanded as  
$A_{\mu}=gA_{\mu}^{a}T_{a}$.  
Here, $g$ is a coupling constant with mass dimension $(2-n)$,  
$\{T_{a} \}$ are Hermitian basis of $\frak{g}$ satisfying 
the commutation relations $[ T_{a}, T_{b} ]=if_{ab}{}^{c} T_{c}$ and the normalization 
conditions ${\rm Tr} (T_{a} T_{b})=\delta_{ab}$.

The Chern-Pontryagin density, $\mathcal{P}_{2n}$, and the Chern-Simons current, 
$\mathcal{C}_{2n}^{\mu}$, on $\mathbf{M}^{2n}$ are essential 
to the $2n$-dimensional models with topological mass generation. 
The Chern-Pontryagin density $\mathcal{P}_{2n}$ is defined by 
\begin{align}
\mathcal{P}_{2n} \equiv\frac{1}{2^{n}} g^n 
h_{a_{1}\cdots a_{n}}  
\epsilon^{\mu_{1}\mu_{2}\cdots\mu_{2n-1}\mu_{2n}} 
F_{\mu_{1} \mu_{2}}^{a_{1}}  \cdots F_{\mu_{2n-1} \mu_{2n}}^{a_{n}}\, , 
\label{1.1}
\end{align}
where $h_{a_{1}\cdots a_{n}}\equiv {\rm Tr}(T_{a_1}\cdots T_{a_{n}})$, and  
$F_{\mu\nu}^{a}$ is the field strength of $A_{\mu}^{a}\,$: 
$F_{\mu\nu}^{a}=\partial_{\mu} A_{\nu}^{a}-\partial_{\nu} A_{\mu}^{a} 
+g f_{bc}{}^{a} A_{\mu}^{b} A_{\nu}^{c}$. 
The Chern-Simons current $\mathcal{C}_{2n}^{\mu}$ is related to $\mathcal{P}_{2n}$  
as follows:   
\begin{align}
\mathcal{P}_{2n} =\partial_{\mu} \mathcal{C}_{2n}^{\mu}\, . 
\label{1.2}
\end{align}
The existence of the Chern-Simons current is guaranteed by Poincar\'{e}'s lemma.

Under the (infinitesimal) gauge transformation 
\begin{align}
\delta_{\omega} A_{\mu}^{a}  = D_{\mu} \omega^{a} , 
\label{1.3}
\end{align}
with $D_{\mu} \omega^{a} \equiv \partial_{\mu} \omega^{a} 
+g f_{bc}{}^{a} A_{\mu}{}^{b} \omega^{c}$,  
$\mathcal{P}_{2n}$ remains invariant, while $\mathcal{C}_{2n}^{\mu}$  
transforms as 
\begin{align}
\delta_{\omega} \mathcal{C}_{2n}^{\nu} 
=\partial_{\mu} \mathcal{U}_{2n}^{\mu\nu} \,. 
\label{1.4}
\end{align}
Here, $\mathcal{U}_{2n}^{\mu\nu}$ is an antisymmetric tensor that is 
a polynomial in $(A_{\mu}^{a}, F_{\mu\nu}^{a}, \omega^{a})$ and linear 
in $\omega^{a}$. The variation of $\mathcal{C}_{2n}^{\mu}$ is found to be  
\begin{align}
\delta \mathcal{C}_{2n}^{\nu} 
&=\mathcal{W}_{2n,a}^{\mu\nu} \delta A^{a}_{\mu} 
+\partial_{\mu} \mathcal{V}_{2n}^{\mu\nu} \, , 
\label{1.5}
\end{align}
where 
\begin{align}
\mathcal{W}_{2n,a}^{\mu\nu} &\equiv 
\frac{n}{2^{n-1}}  g^{n} h
_{a_{1}\cdots a_{n-1} a}  
\epsilon^{\mu_{1}\mu_{2}\cdots \mu_{2n-3} \mu_{2n-2} \mu\nu} 
\nonumber 
\\ 
& \quad \; \times 
F_{\mu_{1} \mu_{2}}^{a_{1}}  \cdots F_{\mu_{2n-3} \mu_{2n-2}}^{a_{n-1}} \,, 
\label{1.6}
\end{align}
and $\mathcal{V}_{2n}^{\mu\nu}$ is an antisymmetric tensor that is 
a polynomial in $(A_{\mu}^{a}, F_{\mu\nu}^{a}, \delta A^{a}_{\mu})$ and linear 
in $\delta A^{a}_{\mu}$. 
(For further details of Eqs. (\ref{1.4}) and (\ref{1.5}),  
see the Appendix of Ref.~\onlinecite{DH}.)

The models with topological mass generation are constructed from  
the topological entities $\mathcal{P}_{2n}$ and $\mathcal{C}_{2n}^{\mu}$ 
and some additional fields and currents \cite{DJP,DH}. 
Among these models, the one that we have called the modified hybrid model 
is self-contained in the sense that the chiral anomaly is incorporated in the model. 
The modified hybrid model is governed by the Lagrangian \cite{DH} 
\begin{align}
\mathcal{L}_{2n}^{\rm top} &= \frac{1}{2} \mathcal{P}_{2n}^{2} 
-\frac{1}{2} m^{2} (\mathcal{C}_{2n}^{\nu} -\partial_{\mu} p^{\mu\nu}) 
(\mathcal{C}_{2n,\nu} -\partial^{\rho} p_{\rho\nu}) 
\nonumber 
\\
& \quad \,
-M\eta \mathcal{P}_{2n} 
+\frac{1}{2} \partial_{\mu} \eta \partial^{\mu} \eta \,, 
\label{1.7} 
\end{align}
where $m$ and $M$ are constants with mass dimension, $p^{\mu\nu}$ is 
an antisymmetric pseudotensor field, and $\eta$ is a pseudoscalar field. 
(The Lagrangian (35) in Ref.~\onlinecite{DH} is reproduced by the replacement 
$M \mapsto \sqrt{N} \Lambda$, $\eta \mapsto \eta_0$.) 
If $m=0$, $\mathcal{L}_{2n}^{\rm top}$ reduces to the $2n$-dimensional generalization 
of a Lagrangian proposed by Dvali {\it et al}. \cite{DJP}. 
If $M=0$, $\mathcal{L}_{2n}^{\rm top}$ is identical to the Lagrangian of 
the St\"{u}ckelberg-type model accompanied by a massless pseudoscalar field  
$\eta$ \cite{DH}. The gauge transformation rules 
\begin{subequations}
\label{1.8}
\begin{align}
\delta_{\omega} p^{\mu\nu} &= \mathcal{U}_{2n}^{\mu\nu} \,, 
\label{1.8a} 
\\
\delta_{\omega} \eta &=0 
\label{1.8b}
\end{align}
\end{subequations}
are imposed on $p^{\mu\nu}$ and $\eta$ so that the Lagrangian 
$\mathcal{L}_{2n}^{\rm top}$ can be gauge invariant. 
Using Eq. (\ref{1.5}), variation of  
the action $S_{2n}^{\rm top}=\int \mathcal{L}_{2n}^{\rm top} dx$ 
with respect to $A^{a}_{\mu}$ is readily calculated,  
yielding the equation of motion 
\begin{align}
& \{ \partial_{\mu} (\mathcal{P}_{2n} -M\eta)
+m^2 (\mathcal{C}_{2n,\mu} -\partial^{\rho} p_{\rho\mu} )  \} 
\mathcal{W}_{2n,a}^{\sigma\mu} 
\nonumber 
\\ 
& 
-m^2 \partial_{\mu} (\mathcal{C}_{2n,\nu} -\partial^{\rho} p_{\rho\nu} ) 
\frac{\delta \mathcal{V}^{\mu\nu}_{2n}}{\delta A^{a}_{\sigma}} =0 \,. 
\label{1.9}
\end{align} 
Variation of $S_{2n}$ with respect to $p^{\mu\nu}$ and $\eta$ yields    
the Euler-Lagrange equations 
\begin{subequations}
\label{1.10}
\begin{align}
&\partial_{\mu} (\mathcal{C}_{2n, \nu} -\partial^{\rho} p_{\rho\nu} ) 
-\partial_{\nu} (\mathcal{C}_{2n, \mu} -\partial^{\rho} p_{\rho\mu} ) =0 \,, 
\label{1.10a}
\\ 
&\sq \eta +M \mathcal{P}_{2n}=0\,, 
\label{1.10b}
\end{align}
\end{subequations}
where $\sq \equiv \partial_{\mu} \partial^{\mu}$. 
By virtue of  Eq. (\ref{1.10a}), the second line of  Eq. (\ref{1.9}) vanishes.  
Also, we can strip away $\mathcal{W}_{2n,a}^{\sigma\mu}$ in Eq. (\ref{1.9}) 
using the identity  
$\mathcal{W}_{2n,a}^{\sigma\mu} F_{\sigma\nu}^a =2\delta^{\mu}_{\nu} \mathcal{P}_{2n}$. 
As a result, provided $\mathcal{P}_{2n} \neq 0$, Eq. (\ref{1.9}) reduces to   
\begin{align}
\partial_{\mu} (\mathcal{P}_{2n} -M \eta) 
+m^2 (\mathcal{C}_{2n,\mu} -\partial^{\nu} p_{\nu\mu} ) =0 \,.
\label{1.11}
\end{align}
Conversely, Eq. (\ref{1.9}) can be reproduced from Eqs. (\ref{1.10a}) and (\ref{1.11}). 
For this reason, it is concluded that 
Eq. (\ref{1.11}) is equivalent to Eq. (\ref{1.9}) with the aid of Eq. (\ref{1.10a}).

Now we consider the axial vector current defined by 
$\mathcal{J}_{\mu}^{5} \equiv\partial_{\mu} \eta$. 
In terms of $\mathcal{J}_{\mu}^{5}$, Eq. (\ref{1.10b}) can be written as 
$\partial^{\mu} \mathcal{J}_{\mu}^{5}=-M\mathcal{P}_{2n}$. 
This shows that the current $\mathcal{J}_{\mu}^{5}$ is not conserved 
due to an anomalous divergence.  
(If $M=0$,  $\mathcal{J}_{\mu}^{5}$ is conserved.) 
In this sense, the modified hybrid model involves its own chiral anomaly without setting  
extra conditions, and consequently 
is recognized as a self-contained model with the chiral anomaly.

It should be noted that Eq. (\ref{1.11}) follows immediately from 
varying $\mathcal{C}_{2n}^{\mu}$, rather than $A^{a}_{\mu}$, in the action 
$S_{2n}^{\rm top}$. 
That is, Eq. (\ref{1.11}) can be derived from $S_{2n}^{\rm top}$ without 
passing through Eq. (\ref{1.9}). 
In this simple way of deriving Eq. (\ref{1.11}), $\mathcal{C}_{2n}^{\mu}$ is treated as 
a fundamental field; it is not necessary to consider the concrete form of 
$\mathcal{C}_{2n}^{\mu}$ written in terms of $A^{a}_{\mu}$. 
The only relation crucial for the simple derivation 
is Eq. (\ref{1.2}). From this fact, we see that 
the modified hybrid model is equivalent to the model governed by the Lagrangian 
\begin{align}
\mathcal{L}_{2n}&= \frac{1}{2} (\partial_{\mu} K^{\mu})^{2} 
-\frac{1}{2} m^{2} (K^{\nu} -\partial_{\mu} p^{\mu\nu}) 
(K_{\nu} -\partial^{\rho} p_{\rho\nu}) 
\nonumber 
\\
& \quad \,
-M  \eta \partial_{\mu} K^{\mu}
+\frac{1}{2} \partial_{\mu} \eta \partial^{\mu} \eta \,. 
\label{1.12} 
\end{align}
Here, $K^{\mu}$ is understood as a fundamental pseudovector field on $\mathbf{M}^{2n}$ 
with no constituents such as $A^{a}_{\mu}$. 
If $K^\mu$ is identified with $\mathcal{C}_{2n}^{\mu}$,  
Eq. (\ref{1.12}) becomes Eq. (\ref{1.7}) by means of Eq. (\ref{1.2}). 
The Lagrangian $\mathcal{L}_{2n}$ is left invariant under the gauge transformation 
\begin{subequations}
\label{2.1}
\begin{align}
\delta_{\lambda} K^{\nu}&=\partial_{\mu} \lambda^{\mu\nu} ,
\label{2.1a}
\\
\delta_{\lambda} p^{\mu\nu}&=\lambda^{\mu\nu} ,
\label{2.1b}
\\
\delta_{\lambda} \eta &=0 \,,
\label{2.1c}
\end{align}
\end{subequations}
where $\lambda^{\mu\nu}$ is a pseudotensorial gauge parameter with 
the antisymmetric property $\lambda^{\mu\nu}=-\lambda^{\nu\mu}$.

The Lagrangian  $\mathcal{L}_{2n}$ can be rewritten as  
\begin{align}
\mathcal{L}_{2n}^{\prime}&=
-\frac{1}{2} P^2 -\frac{1}{2} m^{2} (K^{\nu} -\partial_{\mu} p^{\mu\nu}) 
(K_{\nu} -\partial^{\rho} p_{\rho\nu}) 
\nonumber 
\\ 
& \quad \,
-K^{\mu} \partial_{\mu} (P -M\eta )
+\frac{1}{2} \partial_{\mu} \eta \partial^{\mu} \eta 
\label{2.2} 
\end{align}
up to a total derivative term.  
Here, $P$ is an auxiliary pseudoscalar field satisfying  
\begin{align}
\delta_{\lambda} P =0 \,. 
\label{2.3}
\end{align}
Under the gauge transformation $\delta_{\lambda}$, 
the Lagrangian $\mathcal{L}_{2n}^{\prime}$ remains invariant  
up to a total derivative. 
The equivalence between $\mathcal{L}_{2n}$ and $\mathcal{L}_{2n}^{\prime}$  
can be shown via the use of the field equation 
\begin{align}
P=\partial_{\mu} K^{\mu}
\label{2.4}
\end{align}
or via the path integration over $P$ in the generating functional with 
the Lagrangian $\mathcal{L}_{2n}^{\prime}$. 
(Equation (\ref{2.4}) corresponds to Eq. (\ref{1.2}); 
if $K^{\mu}=\mathcal{C}_{2n}^{\mu}$, it follows that 
$P=\mathcal{P}_{2n}$.)

\section{\label{sec:level1}BRST symmetry and a gauge-fixing term}

In this section, we consider a gauge-fixing procedure aiming at studying quantum-theoretical 
aspects of the equivalent model governed by the Lagrangian $\mathcal{L}_{2n}^{\prime}$. 
For the sake of convenience in later studies,  
we apply the BRST gauge-fixing procedure \cite{KU,NO} 
to the equivalent model, although it is Abelian. 
To this end, we introduce the FP ghost field $C^{\mu\nu}$, 
the FP anti-ghost field 
$\bar{C}^{\mu\nu}$, and the Nakanishi-Lautrup field $B^{\mu\nu}$,  
all of which are assumed to be antisymmetric pseudotensor fields on $\mathbf{M}^{2n}$. 
It is also assumed that $C^{\mu\nu}$ and $\bar{C}^{\mu\nu}$ are anticommutative fields,  
while the other fields are commutative fields. 
The BRST transformation rules of $K^{\mu}$ and $p^{\mu\nu}$ are 
defined by replacing $\lambda^{\mu\nu}$ in Eqs. (\ref{2.1}) by $C^{\mu\nu}$, 
while $\eta$ and $P$ are assumed to be BRST invariant in accordance with 
Eqs. (\ref{2.1c}) and (\ref{2.3}): 
\begin{subequations}
\label{2.5}
\begin{align}
\boldsymbol{\delta} K^{\nu}&=\partial_{\mu} C^{\mu\nu} ,
\label{2.5a}
\\
\boldsymbol{\delta} p^{\mu\nu}&= C^{\mu\nu} ,
\label{2.5b}
\\
\boldsymbol{\delta} \eta &=0 \,,
\label{2.5c}
\\ 
\boldsymbol{\delta} P &=0 \,. 
\label{2.5d}
\end{align}
\end{subequations}
The BRST transformation rules of $C^{\mu\nu}$, $\bar{C}^{\mu\nu}$, and $B^{\mu\nu}$ are 
defined by 
\begin{subequations}
\label{2.6}
\begin{align}
\boldsymbol{\delta} C^{\mu\nu}&=0 \,,
\label{2.6a}
\\
\boldsymbol{\delta} \bar{C}^{\mu\nu}&=iB^{\mu\nu} ,
\label{2.6b}
\\
\boldsymbol{\delta} B^{\mu\nu}&=0 \,,
\label{2.6c}
\end{align}
\end{subequations}
in such a way that 
the nilpotency property $\boldsymbol{\delta}^2 =0$ is valid for all the fields.

Now we adopt the gauge-fixing (GF) condition 
$K_{\mu\nu}-\alpha B_{\mu\nu}=0$ 
in order that the gauge invariance of $\mathcal{L}_{2n}^{\prime}$ 
(up to a total derivative) can be broken. 
Here, $K_{\mu\nu}\equiv \partial_{\mu} K_{\nu} -\partial_{\nu} K_{\mu}\,$, 
and $\alpha$ is a gauge parameter. 
In the BRST gauge-fixing procedure, the condition $K_{\mu\nu}-\alpha B_{\mu\nu}=0$  
is incorporated in the sum of gauge-fixing and FP ghost terms  
(or simply the gauge-fixing term) 
\begin{align}
\mathcal{L}_{\rm GF} =-\frac{i}{2} \boldsymbol{\delta} 
\bigg[\, \bar{C}^{\mu\nu} \bigg( K_{\mu\nu} -\frac{\alpha}{2} B_{\mu\nu} \bigg) \bigg] \,. 
\label{2.7}
\end{align}
The BRST invariance of $\mathcal{L}_{\rm GF}$ is guaranteed by 
the nilpotency of $\boldsymbol{\delta}$. 
In contrast, the BRST invariance of $\mathcal{L}_{2n}^{\prime}$ (up to a total derivative) 
is clear from its gauge invariance (up to a total derivative).  
Carrying out the BRST transformation contained in the right-hand side of Eq. (\ref{2.7}) 
and adding the resultant to Eq. (\ref{2.2}), we have the (total) Lagrangian 
\begin{align}
\hat{\mathcal{L}}_{2n}&= 
-\frac{1}{2} P^2 -\frac{1}{2} m^{2} (K^{\nu} -\partial_{\mu} p^{\mu\nu}) 
(K_{\nu} -\partial^{\rho} p_{\rho\nu}) 
\nonumber 
\\ 
& \quad \,
-K^{\mu} \partial_{\mu} (P -M\eta )
+\frac{1}{2} \partial_{\mu} \eta \partial^{\mu} \eta 
\nonumber 
\\ 
& \quad \, 
-K^{\mu} \partial^{\nu} B_{\nu\mu} -{\alpha\over 4} B_{\mu\nu} B^{\mu\nu}
-i \partial_{\mu} \bar{C}^{\mu\nu} \partial^{\rho} C_{\rho\nu} \,.
\label{2.8} 
\end{align}
Here, a total derivative has been removed.

From the Lagrangian $\hat{\mathcal{L}}_{2n}$, 
the Euler-Lagrange equations for $K^{\mu}$, $p^{\mu\nu}$, $\eta$, $P$, 
$B^{\mu\nu}$, $\bar{C}^{\mu\nu}$, and $C^{\mu\nu}$ are derived, respectively, as 
\begin{subequations}
\label{2.9}
\begin{align}
&\partial_{\mu} (P-M\eta) +m^2 (K_\mu -\partial^{\nu} p_{\nu\mu})
+\partial^{\nu} B_{\nu\mu} =0 \,, 
\label{2.9a}
\\
&\partial_{\mu}(K_{\nu}-\partial^{\rho} p_{\rho\nu}) 
-\partial_{\nu}(K_{\mu}-\partial^{\rho} p_{\rho\mu}) =0 \,, 
\label{2.9b}
\\
&\sq \eta +M \partial_{\mu} K^{\mu} =0 \,, 
\label{2.9c}
\\
&\partial_{\mu} K^{\mu} -P =0 \,, 
\label{2.9d}
\\
&\partial_{\mu} K_{\nu} -\partial_{\nu} K_{\mu} -\alpha B_{\mu\nu} =0 \,, 
\label{2.9e}
\\
&\partial_{\mu} \partial^{\rho} C_{\rho\nu} -\partial_{\nu} \partial^{\rho} C_{\rho\mu} =0 \,, 
\label{2.9f}
\\
&\partial_{\mu} \partial^{\rho} \bar{C}_{\rho\nu} 
-\partial_{\nu} \partial^{\rho} \bar{C}_{\rho\mu} =0 \,, 
\label{2.9g}
\end{align}
\end{subequations}
where $\sq \equiv \partial_{\mu} \partial^{\mu}$. 
Using Eqs. (\ref{2.9a}) and (\ref{2.9b}), we can show that 
\begin{align}
\partial_{\mu} \partial^{\rho} B_{\rho\nu} -\partial_{\nu} \partial^{\rho} B_{\rho\mu} =0 \,. 
\label{2.10}
\end{align}
This can also be derived from the BRST transformation of Eq. (\ref{2.9g}).  
Combining Eqs. (\ref{2.9b}) and (\ref{2.9e}) gives    
\begin{align}
\partial_{\mu} \partial^{\rho} p_{\rho\nu} -\partial_{\nu} \partial^{\rho} p_{\rho\mu} 
-\alpha B_{\mu\nu}=0 \,.
\label{2.11}
\end{align}
With Eq. (\ref{2.9d}), Eq. (\ref{2.9c}) reads 
\begin{align}
\sq \eta +M P =0 \,. 
\label{2.12}
\end{align}
Taking the divergence of Eq. (\ref{2.9a}) and using Eqs. (\ref{2.9d}) and (\ref{2.12}), 
we obtain, due to antisymmetry of $p_{\nu\mu}$ and $B_{\nu\mu}$ in their indices, 
\begin{align}
(\sq +M^{2}+m^{2}) P =0 \,. 
\label{2.13}
\end{align}
Taking the divergence of Eq. (\ref{2.9e}) gives  
\begin{align}
\sq K_{\mu} -\partial_{\mu} P -\alpha \partial^{\nu} B_{\nu\mu}=0 \,, 
\label{2.14}
\end{align}
which, together with Eq. (\ref{2.10}),  leads to 
\begin{align}
\sq K_{\mu\nu}=0 \,. 
\label{2.15}
\end{align}
Taking the divergence of Eqs. (\ref{2.11}), (\ref{2.10}), (\ref{2.9f}) and (\ref{2.9g}) 
yields 
\begin{subequations}
\label{2.16}
\begin{align}
& \sq \partial^{\rho} p_{\rho\nu} -\alpha \partial^{\mu} B_{\mu\nu} =0 \,, 
\label{2.16a}
\\ 
& \sq \partial^{\rho} B_{\rho\nu} =0 \,, 
\label{2.16b}
\\
& \sq \partial^{\rho} C_{\rho\nu} =0 \,, 
\label{2.16c}
\\ 
& \sq \partial^{\rho} \bar{C}_{\rho\nu} =0 \,. 
\label{2.16d}
\end{align}
\end{subequations}
In the remaining sections, we mainly study the canonical 
formalism of the equivalent model  
and its application to the quantization of this model. 
Based on this study, we investigate particle contents of 
the equivalent model and of the modified hybrid model.

\section{\label{sec:level1}Canonical formalism} 

For a while, we treat the fields introduced above 
as canonical coordinates and collectively express them as  
$(\varPhi^{I})\equiv(K^{\mu}, P, \,\eta, \,p^{\mu\nu}, B^{\mu\nu}, C^{\mu\nu}, \bar{C}^{\mu\nu})$, 
where $I$ stands for space-time indices.  With the Lagrangian (\ref{2.8}), 
the canonical momentum conjugate to $\varPhi^{I}$ is defined by 
\begin{align}
\varPi^{\varPhi}_{I} \equiv \frac{\partial \hat{\mathcal{L}}_{2n}}{\partial \dot{\varPhi}^{I}} \,, 
\label{3.1}
\end{align}
where $\dot{f} \equiv \partial f/ \partial t$. For the anticommutative fields 
$C^{\mu\nu}$ and $\bar{C}^{\mu\nu}$, the derivative in Eq. (\ref{3.1}) is understood as  
the left derivative. We can readily find 
\begin{subequations}
\label{3.2}
\begin{align}
\varPi^{K}_{\mu}&=0 \,, 
\label{3.2a}
\\
\varPi^{P}&=-K_{0} \,, 
\label{3.2b}
\\
\varPi^{\eta}&=MK_{0} +\dot{\eta} \,, 
\label{3.2c}
\\
\varPi^{p}_{0j}&=-m^{2} (\dot{p}_{0j}-\partial_{i} p_{ij} -K_{j}) \,, 
\label{3.2d}
\\
\varPi^{p}_{ij}&=0 \,, 
\label{3.2e}
\\
\varPi^{B}_{0j}&=-K_{j} \,, 
\label{3.2f}
\\
\varPi^{B}_{ij}&=0 \,, 
\label{3.2g}
\\
\varPi^{C}_{0j}&=i(\dot{\bar{C}}_{0j} -\partial_{i} \bar{C}_{ij}) \,, 
\label{3.2h}
\\
\varPi^{C}_{ij}&=0 \,, 
\label{3.2i}
\\
\varPi^{\bar{C}}_{0j}&=-i(\dot{C}_{0j} -\partial_{i} C_{ij}) \,, 
\label{3.2j}
\\
\varPi^{\bar{C}}_{ij}&=0 \,. 
\label{3.2k}
\end{align}
\end{subequations}
The Hamiltonian density is obtained from Eqs. (\ref{2.8}) and (\ref{3.2}): 
\begin{align}
\mathcal{H}_{2n} &\equiv 
\dot{K}^{\mu} \varPi^{K}_{\mu} +\dot{P} \varPi^{P} +\dot{\eta} \varPi^{\eta}
+\frac{1}{2} (\dot{p}^{\mu\nu} \varPi^{p}_{\mu\nu} 
+\dot{B}^{\mu\nu} \varPi^{B}_{\mu\nu} 
+\dot{C}^{\mu\nu} \varPi^{C}_{\mu\nu} 
+\dot{\bar{C}}^{\mu\nu} \varPi^{\bar{C}}_{\mu\nu} ) -\hat{\mathcal{L}}_{2n} 
\nonumber 
\\
& =\frac{1}{2} (M^2 +m^2) (\varPi^{P})^{2} +\frac{1}{2} P^{2} +\frac{1}{2} (\varPi^{\eta})^{2} 
+\frac{1}{2} \partial_{j}\eta \partial_{j}\eta +M\varPi^{P} \varPi^{\eta}
+\varPi^{P} \partial_{j} (m^2 p_{0j}-B_{0j}) 
\nonumber 
\\
& \quad\, +\frac{1}{2m^2} \varPi^{p}_{0j} \varPi^{p}_{0j} 
+\frac{1}{2} m^2 \partial_{i} p_{0i} \partial_{j} p_{0j} 
+\varPi^{p}_{0j} \varPi^{B}_{0j} -\varPi^{p}_{0j} \partial_{i} p_{ij} 
+\varPi^{B}_{0j} (\partial_{j} P -M\partial_{j} \eta -\partial_{i} B_{ij}) 
\nonumber 
\\
& \quad\, -\frac{\alpha}{4} (2B_{0j} B_{0j} -B_{ij} B_{ij}) 
-i\varPi^{\bar{C}}_{0j} \varPi^{C}_{0j} 
+i \partial_{i} \bar{C}_{0i} \partial_{j} C_{0j} 
+\varPi^{\bar{C}}_{0j} \partial_{i} \bar{C}_{ij} 
+\varPi^{C}_{0j} \partial_{i} C_{ij} \,. 
\label{3.3}
\end{align}
The Poisson bracket of two arbitrary monomial functions of the canonical variables,    
$F=F(\varPhi^{I}, \varPi^{\varPhi}_{J})$ and $G=G(\varPhi^{I}, \varPi^{\varPhi}_{J})$,  
is defined by 
\begin{align}
\{F, G\}=\int d^{2n-1} \boldsymbol{x} 
\bigg[ (-1)^{|F||\varPhi^{I}|}
\frac{\partial F}{\partial \varPhi^{I}(t,\boldsymbol{x})} 
\frac{\partial G}{\partial \varPi^{\varPhi}_{I}(t,\boldsymbol{x})} 
-(-1)^{|G|(|\varPhi^{I}|+|F|)}
\frac{\partial G}{\partial \varPhi^{I}(t,\boldsymbol{x})} 
\frac{\partial F}{\partial \varPi^{\varPhi}_{I}(t,\boldsymbol{x})} 
\bigg] 
\label{3.4}
\end{align}
in such a way that it reduces to the following Poisson brackets: 
\begin{subequations}
\label{3.5}
\begin{align}
&\{ K^{\mu}(t,\boldsymbol{x}), \varPi^{K}_{\nu}(t,\boldsymbol{y}) \}
=\delta^{\mu}_{\nu} \delta(\boldsymbol{x}-\boldsymbol{y}) \,, 
\label{3.5a}
\\
&\{ P(t,\boldsymbol{x}), \varPi^{P}(t,\boldsymbol{y}) \}
=\delta(\boldsymbol{x}-\boldsymbol{y}) \,, 
\label{3.5b}
\\
&\{ \eta(t,\boldsymbol{x}), \varPi^{\eta}(t,\boldsymbol{y}) \}
=\delta(\boldsymbol{x}-\boldsymbol{y}) \,, 
\label{3.5c}
\\
&\{ p^{\mu\nu}(t,\boldsymbol{x}), \varPi^{p}_{\rho\sigma}(t,\boldsymbol{y}) \}
=\delta^{\mu\nu}_{\rho\sigma} 
\delta(\boldsymbol{x}-\boldsymbol{y}) \,, 
\label{3.5d}
\\
&\{ B^{\mu\nu}(t,\boldsymbol{x}), \varPi^{B}_{\rho\sigma}(t,\boldsymbol{y}) \}
=\delta^{\mu\nu}_{\rho\sigma} 
\delta(\boldsymbol{x}-\boldsymbol{y}) \,, 
\label{3.5e}
\\
&\{ C^{\mu\nu}(t,\boldsymbol{x}), \varPi^{C}_{\rho\sigma}(t,\boldsymbol{y}) \}
=-\delta^{\mu\nu}_{\rho\sigma} 
\delta(\boldsymbol{x}-\boldsymbol{y}) \,, 
\label{3.5f}
\\
&\{ \bar{C}^{\mu\nu}(t,\boldsymbol{x}), \varPi^{\bar{C}}_{\rho\sigma}(t,\boldsymbol{y}) \}
=-\delta^{\mu\nu}_{\rho\sigma} 
\delta(\boldsymbol{x}-\boldsymbol{y}) \,, 
\label{3.5g}
\end{align}
\end{subequations}
where $\delta^{\mu\nu}_{\rho\sigma} \equiv 
\delta^{\mu}_{\rho} \delta^{\nu}_{\sigma}-\delta^{\nu}_{\rho} \delta^{\mu}_{\sigma}$. 
The symbol $|F|$ takes the value 0 or 1  
according as $F$ is an even or odd power with respect to the anticommutative 
canonical variables. 
The Poisson bracket (\ref{3.4}) satisfies $\{F, G\}=-(-1)^{|F||G|} \{G, F\}$.

Equations (\ref{3.2}), 
except Eqs. (\ref{3.2c}), (\ref{3.2d}), (\ref{3.2h}) and (\ref{3.2j}), 
are read as the primary constraints 
\begin{subequations}
\label{3.6}
\begin{align}
\phi^{1} &\equiv \varPi^{K}_{0} \approx 0 \,, 
\label{3.6a}
\\
\phi^{2}_{i} &\equiv \varPi^{K}_{i} \approx 0 \,, 
\label{3.6b}
\\
\phi^{3} & \equiv \varPi^{P}+K_{0} \approx 0 \,, 
\label{3.6c}
\\
\phi^{4}_{0j} & \equiv \varPi^{B}_{0j} +K_{j} \approx 0 \,, 
\label{3.6d}
\\
\phi^{5}_{ij} & \equiv \varPi^{p}_{ij} \approx0 \,, 
\label{3.6e}
\\
\phi^{6}_{ij} & \equiv \varPi^{B}_{ij} \approx0 \,, 
\label{3.6f}
\\
\phi^{7}_{ij} & \equiv \varPi^{C}_{ij} \approx 0 \,, 
\label{3.6g}
\\
\phi^{8}_{ij} & \equiv \varPi^{\bar{C}}_{ij} \approx 0 \,, 
\label{3.6h}
\end{align}
\end{subequations}
where the symbol ^^ ^^ $\approx$" denotes the weak equality. 
Now we apply the Dirac formulation for constrained Hamiltonian systems 
\cite{Dir,HRT,HT} to the present model. 
Introducing the Lagrange multipliers  
$v^{a}_{I}=v^{a}_{I}(t,\boldsymbol{x})$ $(a=1, 2, \cdots ,8)$,  
we define the {\em total} Hamiltonian density 
\begin{align}
\hat{\mathcal{H}}_{2n} & \equiv \mathcal{H}_{2n}
+v^{1} \phi^{1} +v^{2}_{i} \phi^{2}_{i} +v^{3} \phi^{3} +v^{4}_{0j} \phi^{4}_{0j}
\nonumber 
\\ 
& \quad \, 
+\frac{1}{2}( v^{5}_{ij} \phi^{5}_{ij} +v^{6}_{ij} \phi^{6}_{ij} 
+ v^{7}_{ij} \phi^{7}_{ij} +v^{8}_{ij} \phi^{8}_{ij} ) 
\label{3.7}
\end{align}
and the {\em total} Hamiltonian 
\begin{align}
\hat{H}_{2n} \equiv 
\int  d^{2n-1} \boldsymbol{x} \, \hat{\mathcal{H}}_{2n} \,.
\label{3.8}
\end{align}
With this Hamiltonian, the canonical equation for $F$ is given by   
\begin{align}
\dot{F} = \{F, \hat{H}_{2n} \} \,.
\label{3.9}
\end{align}
The primary constraints (\ref{3.6}) must be preserved in time 
so that they can be consistent with the equations of motion. 
Hence, if we take $F$ in Eq. (\ref{3.9}) to be one of $\phi^{a}_{I}$, 
we should have 
$\dot{\phi}^{a}_{I} =\{{\phi}^{a}_{I}, \hat{H}_{2n} \} \approx 0$. 
The consistency conditions $\dot{\phi}^{1} \approx 0$, 
$\dot{\phi}^{2}_{i} \approx 0$, $\dot{\phi}^{3} \approx 0$,  
and $\dot{\phi}^{4}_{0j} \approx 0$ determine the Lagrange multipliers  
$v^{3}$, $v^{4}_{0j}$, $v^{1}$, and $v^{2}_{i}$, respectively, as  
\begin{subequations}
\label{3.10}
\begin{align}
v^{3} &\approx 0 \,, 
\label{3.10a}
\\
v^{4}_{0j} &\approx 0 \,, 
\label{3.10b}
\\
v^{1} &\approx P-\partial_{j} \varPi^{B}_{0j} \,, 
\label{3.10c}
\\
v^{2}_{i} &\approx \partial_{i} \varPi^{P} -\alpha B_{0i} \,. 
\label{3.10d}
\end{align}
\end{subequations} 
The consistency conditions 
$\dot{\phi}^{5}_{ij} \approx 0$,  $\dot{\phi}^{6}_{ij} \approx 0$, 
$\dot{\phi}^{7}_{ij} \approx 0$, and $\dot{\phi}^{8}_{ij} \approx 0$ give rise to  
the secondary constraints   
\begin{subequations}
\label{3.11}
\begin{align}
\phi^{9}_{ij} &\equiv \partial_{i} \varPi^{p}_{0j} -\partial_{j} \varPi^{p}_{0i} \approx 0 \,, 
\label{3.11a}
\\ 
\phi^{10}_{ij} &\equiv \partial_{i} \varPi^{B}_{0j} -\partial_{j} \varPi^{B}_{0i} +\alpha B_{ij} \approx 0 \,, 
\label{3.11b}
\\
\phi^{11}_{ij} &\equiv \partial_{i} \varPi^{C}_{0j} -\partial_{j} \varPi^{C}_{0i} \approx 0 \,, 
\label{3.11c}
\\
\phi^{12}_{ij} &\equiv \partial_{i} \varPi^{\bar{C}}_{0j} -\partial_{j} \varPi^{\bar{C}}_{0i} \approx 0 \,,  
\label{3.11d}
\end{align}
\end{subequations} 
respectively. We can also evaluate the time evolutions of 
$\phi^{9}_{ij}$, $\phi^{11}_{ij}$, and $\phi^{12}_{ij}$ using Eq. (\ref{3.9}), 
and see that the equations 
$\dot{\phi}^{9}_{ij}=0$, $\dot{\phi}^{11}_{ij}=0$, and $\dot{\phi}^{12}_{ij}=0$ 
are identically satisfied.  
For $\phi^{10}_{ij}$, its time evolution is found to be 
\begin{align}
\dot{\phi}^{10}_{ij}=\alpha (v^{6}_{ij} -\partial_{i} B_{0j} +\partial_{j} B_{0i} ) \,.
\label{3.12}
\end{align}
If $\alpha\neq 0$, the condition $\dot{\phi}^{10}_{ij} \approx 0$  
determines the Lagrange multiplier $v^{6}_{ij}$ as 
\begin{align}
v^{6}_{ij} \approx \partial_{i} B_{0j} -\partial_{j} B_{0i} \,. 
\label{3.13}
\end{align}
If $\alpha=0$, $\dot{\phi}^{10}_{ij}$ identically vanishes: $\dot{\phi}^{10}_{ij}=0$. 
In both the cases $\alpha\neq 0$ and $\alpha=0$,  
no further secondary constraints are derived, 
and thus the derivation of constraints is completed at present. 
The constraints that we need to consider are therefore 
$\phi^{\hat{a}}_{I} \approx 0$ $(\hat{a}=1, 2, \cdots ,12)$ 
stated in Eqs. (\ref{3.6}) and (\ref{3.11}).

When $\alpha\neq 0$, using the Poisson brackets (\ref{3.5}), it can be shown that 
the constraints 
$\phi^{1}\approx 0$, $\phi^{2}_{i}\approx 0$, $\phi^{3}\approx 0$, $\phi^{4}_{0j}\approx 0$,  
$\phi^{6}_{ij}\approx 0$, and $\phi^{10}_{ij}\approx 0$ 
are classified into second class, while 
the other six constraints are classified into first class. 
Accordingly, the Lagrange multipliers $v^{1}$, $v^{2}_{i}$, $v^{3}$, $v^{4}_{0j}$, and 
$v^{6}_{ij}$ are determined to be zero or to be  
what is written in terms of the canonical variables, 
as can be seen in Eqs. (\ref{3.10}) and (\ref{3.13}). 
The other multipliers $v^{5}_{ij}$, $v^{7}_{ij}$, and $v^{8}_{ij}$ remain arbitrary. 
When $\alpha= 0$, it can be shown that only the constraints 
$\phi^{1}\approx 0$, $\phi^{2}_{i}\approx 0$, $\phi^{3}\approx 0$, and $\phi^{4}_{0j}\approx 0$ 
are classified into second class, while the other eight constraints are classified into first class. 
Accordingly, only the Lagrange multipliers $v^{1}$, $v^{2}_{i}$, $v^{3}$, and $v^{4}_{0j}$ 
are determined to be zero or to be what is written in terms of the canonical variables; 
the other multipliers $v^{5}_{ij}$, $v^{6}_{ij}$, $v^{7}_{ij}$, and $v^{8}_{ij}$ remain arbitrary. 
As regards a pair of the constraints $\phi^{6}_{ij}\approx 0$ and $\phi^{10}_{ij}\approx 0$, 
its treatment in the case $\alpha\neq 0$ is thus different from that in the case $\alpha\neq 0$.  
In what follows, we consider only the case $\alpha=0$ for the sake of simplicity, 
although the case $\alpha \neq 0$ can be discussed with no difficulties.

Now, we impose the gauge-fixing conditions 
\begin{subequations}
\label{3.14}
\begin{align}
\chi^{1}_{ij} &\equiv  p_{ij} \approx 0 \,, 
\label{3.14a}
\\
\chi^{2}_{ij} &\equiv  B_{ij} \approx 0 \,, 
\label{3.14b}
\\
\chi^{3}_{ij} &\equiv  C_{ij} \approx 0 \,, 
\label{3.14c}
\\
\chi^{4}_{ij} &\equiv  \bar{C}_{ij} \approx 0 \,, 
\label{3.14d}
\end{align}
\end{subequations}
to make the first-class primary constraints Eqs. (\ref{3.6e})-(\ref{3.6h}) second class. 
From Eqs. (\ref{3.5d})-(\ref{3.5g}), it follows that 
$\{ \phi_{ij}^{a^{\prime}+4}(t,\boldsymbol{x}), \chi_{ij}^{a^{\prime}}(t,\boldsymbol{x}) \}
=-\delta(\boldsymbol{0})\neq 0$ (${a^{\prime}}=1, 2, 3, 4$). 
(Here, no summation over $i$ and $j$ is taken.)  
These relations guarantee that Eqs (\ref{3.14}) function as gauge-fixing conditions, 
and thus the $\phi_{ij}^{a^{\prime}+4} \approx 0$ and $\chi_{ij}^{a^{\prime}} \approx 0$ 
are together classified into second class. 
The gauge-fixing conditions must be preserved in time 
so that they can be consistent with the equations of motion; 
hence, we should 
have $\dot{\chi}^{a^{\prime}}_{ij} =\{{\chi}^{a^{\prime}}_{ij}, \hat{H}_{2n} \} \approx 0$. 
These consistency conditions determine the Lagrange multipliers   
$v^{a^{\prime}+4}_{ij}$  (${a^{\prime}}=1, 2, 3, 4$) as 
\begin{align}
v^{a^{\prime}+4}_{ij} \approx 0 \,.
\label{3.15}
\end{align}
Up to here, all the Lagrange multipliers $v^{a}_{I}$ have been determined 
as in Eqs. (\ref{3.10}) and (\ref{3.15}). 
This implies that the gauge degrees of freedom are now completely fixed.

Using $\phi^{a}_{I} \approx 0$ 
$(a=1,2, \cdots 8)$ and  
$\chi_{ij}^{a^{\prime}} \approx 0$, which constitute 
second-class constraints, 
we define the Dirac bracket:   
\begin{align}
\{F, G\}_{\rm D}
&= \{F, G\} -\int d^{2n-1} \boldsymbol{x} 
\bigg[ \{F, \phi^{1}(t,\boldsymbol{x}) \} \{ \phi^{3}(t,\boldsymbol{x}), G\} 
-\{F, \phi^{2}_{i} (t,\boldsymbol{x}) \} \{ \phi^{4}_{0i} (t,\boldsymbol{x}), G\} 
\nonumber 
\\
& \quad\, -\frac{1}{2} \sum_{a^{\prime}=1}^4 
\{F, \chi^{a^{\prime}}_{ij} (t,\boldsymbol{x}) \} 
\{  \phi^{a^{\prime}+4}_{ij} (t,\boldsymbol{x}), G\} 
-(-1)^{|F||G|} (F \leftrightarrow G) \bigg] \,. 
\label{3.16}
\end{align}
Because $\{F, \phi^{a}_{I} \}_{\rm D}=\{F, \chi_{ij}^{a^{\prime}} \}_{\rm D}=0$  
is valid for any $F$, the primary constraints (\ref{3.6}) and 
the gauge-fixing conditions (\ref{3.14}) can be set equal to zero even 
before evaluating Dirac brackets.  
That is, with the Dirac bracket (\ref{3.16}), Eqs. (\ref{3.6}) and (\ref{3.14}) 
can be treated as {\em strong} equations, and  
may be expressed as $\phi^{a}_{I} =0$ and $\chi_{ij}^{a^{\prime}} =0$. 
From the Hamiltonian density (\ref{3.3}), we define the {\em reduced} Hamiltonian 
\begin{align}
\tilde{H}_{2n}
&\equiv 
\int  d^{2n-1} \boldsymbol{x}  \, 
\mathcal{H}_{2n}(\chi_{ij}^{a^{\prime}} =0, \alpha=0) 
\nonumber 
\\
&=\int  d^{2n-1} \boldsymbol{x}  \bigg[ \,
\frac{1}{2} (M^2 +m^2) (\varPi^{P})^{2} +\frac{1}{2} P^{2} +\frac{1}{2} (\varPi^{\eta})^{2} 
+\frac{1}{2} \partial_{j}\eta \partial_{j}\eta +M\varPi^{P} \varPi^{\eta}
\nonumber 
\\
& \quad\, +\varPi^{P} \partial_{j} (m^2 p_{0j}-B_{0j}) 
+\frac{1}{2m^2} \varPi^{p}_{0j} \varPi^{p}_{0j} 
+\frac{1}{2} m^2 \partial_{i} p_{0i} \partial_{j} p_{0j} 
+\varPi^{p}_{0j} \varPi^{B}_{0j} 
\nonumber 
\\
& \quad\, 
+\varPi^{B}_{0j} (\partial_{j} P -M\partial_{j} \eta ) 
-i\varPi^{\bar{C}}_{0j} \varPi^{C}_{0j} 
+i \partial_{i} \bar{C}_{0i} \partial_{j} C_{0j} \bigg] \,. 
\label{3.17}
\end{align}
Owing to the consistency conditions
$\dot{\phi}^{a}_{I} \approx 0$ and $\dot{\chi}_{ij}^{a^{\prime}} \approx 0$, 
the weak equality 
$\{F, \hat{H}_{2n} \}\approx \{F, \hat{H}_{2n} \}_{\rm D}$  is valid for any $F$. 
Using this equality and $\{F, \hat{H}_{2n} \}_{\rm D}=\{F, \tilde{H}_{2n} \}_{\rm D}$, 
the canonical equation (\ref{3.9}) can be written    
\begin{align}
\dot{F} \approx \{F, \tilde{H}_{2n} \}_{\rm D} \,.
\label{3.18}
\end{align}

So far the secondary constraints (\ref{3.11}) have been left first class. 
Because the primary constraints (\ref{3.6}) are now treated as strong equations 
by virtue of the Dirac bracket, 
the weak equalities in Eqs. (\ref{3.11}) should be reconsidered as strong equalities,  
with replacing the symbol ^^ ^^ $\approx$" by ^^ ^^ $=$". 
Noting this fact, we solve Eqs. (\ref{3.11}), including Eq. (\ref{3.11b}) with $\alpha=0$, 
in terms of pseudoscalar functions in the sense of strong equations: 
\begin{subequations}
\label{3.19}
\begin{align}
\varPi^{p}_{0j} & =\partial_{j} \varPi^{p} ,
\label{3.19a}
\\
\varPi^{B}_{0j} & =\partial_{j} \varPi^{B} ,
\label{3.19b}
\\
\varPi^{C}_{0j} &=\partial_{j} \varPi^{C} ,
\label{3.19c}
\\
\varPi^{\bar{C}}_{0j} &=\partial_{j} \varPi^{\bar{C}} ,
\label{3.19d}
\end{align}
\end{subequations}
where $\varPi^{p}$ and $\varPi^{B}$ are commutative functions, while 
$\varPi^{C}$, and $\varPi^{\bar{C}}$ are anticommutative functions. 
Equations (\ref{3.19}) are valid at least in a local region of the phase space. 
In this way, the secondary constraints (\ref{3.11}) have completely been solved 
in terms of the pseudoscalar functions $\varPi^{p}$, $\varPi^{B}$, $\varPi^{C}$,  
and $\varPi^{\bar{C}}$, 
and consequently we do not need to consider gauge-fixing conditions for 
these constraints.

Now, consider the Poisson bracket 
$\{ p_{0i}(t,\boldsymbol{x}), \varPi^{p}_{0j}(t,\boldsymbol{y}) \}
=-\delta_{ij}\delta(\boldsymbol{x}-\boldsymbol{y})$ given from Eq. (\ref{3.5d}). 
Because of 
$\{ \phi^{a}_{I} (t,\boldsymbol{x}), \varPi^{p}_{0j}(t,\boldsymbol{y}) \}
= \{ \chi_{ik}^{a^{\prime}} (t,\boldsymbol{x}), \varPi^{p}_{0j}(t,\boldsymbol{y}) \}=0$, 
the corresponding Dirac bracket takes the same form: 
$\{ p_{0i}(t,\boldsymbol{x}), \varPi^{p}_{0j}(t,\boldsymbol{y}) \}_{\rm D}
=-\delta_{ij}\delta(\boldsymbol{x}-\boldsymbol{y})$.  
Substituting Eq. (\ref{3.19a}) into this bracket and taking the divergence 
of $p_{0i}$ in the bracket, we have       
\begin{align}
\frac{\partial}{\partial y^j }
\{ p (t,\boldsymbol{x}), 
\varPi^{p} (t,\boldsymbol{y}) \}_{\rm D} 
=\frac{\partial}{\partial y^j } \delta(\boldsymbol{x}-\boldsymbol{y}) \,, 
\label{3.20}
\end{align}
with  
\begin{align}
p \equiv \partial_{i} p_{0i} \,. 
\label{3.21}
\end{align}
Integrating Eq. (\ref{3.20}) over $(y^j)$ leads to     
$ \{ p (t,\boldsymbol{x}),  
\varPi^{p} (t,\boldsymbol{y}) \}_{\rm D}  
=\delta(\boldsymbol{x}-\boldsymbol{y}) +f (t, \boldsymbol{x})$, 
where $f$ is a smooth function on Minkowski space $\mathbf{M}^{2n}$.  
To maintain the locality in the system, we must set the condition $f=0$, 
and hence obtain    
\begin{align}
\{ p (t,\boldsymbol{x}), 
\varPi^{p} (t,\boldsymbol{y}) \}_{\rm D}
=\delta(\boldsymbol{x}-\boldsymbol{y}) \,. 
\label{3.22}
\end{align}
Following the procedure used in deriving Eq. (\ref{3.22}) from Eq. (\ref{3.5d}), 
we can derive from Eqs. (\ref{3.5e})-(\ref{3.5g}) the following Dirac brackets:  
\begin{subequations}
\label{3.23}
\begin{align}
& \{ B (t,\boldsymbol{x}), 
\varPi^{B} (t,\boldsymbol{y}) \}_{\rm D} 
=\delta(\boldsymbol{x}-\boldsymbol{y}) \,, 
\label{3.23a}
\\
& \{ C (t,\boldsymbol{x}), 
\varPi^{C} (t,\boldsymbol{y}) \}_{\rm D} 
=-\delta(\boldsymbol{x}-\boldsymbol{y}) \,, 
\label{3.23b}
\\
& \{ \bar{C} (t,\boldsymbol{x}), 
\varPi^{\bar{C}} (t,\boldsymbol{y}) \}_{\rm D} 
=-\delta(\boldsymbol{x}-\boldsymbol{y}) \,, 
\label{3.23c}
\end{align}
\end{subequations}
with  
\begin{subequations}
\label{3.24}
\begin{align}
B \equiv \partial_{i} B_{0i} \,, 
\label{3.24a}
\\
C \equiv \partial_{i} C_{0i} \,, 
\label{3.24b}
\\
\bar{C} \equiv \partial_{i} \bar{C}_{0i} \,. 
\label{3.24c}
\end{align}
\end{subequations}
Having obtained Eqs. (\ref{3.22}) and (\ref{3.23}), 
we can regard $p$, $B$, $C$, and $\bar{C}$ as canonical coordinates, while 
$\varPi^{p}$, $\varPi^{B}$, $\varPi^{C}$, and $\varPi^{\bar{C}}$ as the momenta 
conjugate to $p$, $B$, $C$, and $\bar{C}$, respectively.

With the Dirac bracket (\ref{3.16}), it is sufficient to consider only 
the pseudoscalar fields 
$P$, $\eta$, $p$, $B$, $C$, and $\bar{C}$ as canonical coordinates. 
We collectively express them as  
$(\varPsi)\equiv(P, \eta, p, B, C, \bar{C})$. 
The canonical momenta conjugate to $(\varPsi)$ are collected to be 
$(\varPi^{\varPsi})=(\varPi^{P}, \varPi^{\eta}, 
\varPi^{p}, \varPi^{B}, \varPi^{C}, \varPi^{\bar{C}})$. 
On the phase space submanifold, $\mathcal{S}$, 
with local coordinates $(\varPsi, \varPi^{\varPsi})$, 
the Dirac bracket (\ref{3.16}) is equivalent to the {\em modified} Poisson bracket 
\begin{align}
\{F, G\}^{\ast}=\int d^{2n-1} \boldsymbol{x} 
\bigg[ (-1)^{|F||\varPsi|}
\frac{\partial F}{\partial \varPsi(t,\boldsymbol{x})} 
\frac{\partial G}{\partial \varPi^{\varPsi}(t,\boldsymbol{x})} 
-(-1)^{|G|(|\varPsi|+|F|)}
\frac{\partial G}{\partial \varPsi(t,\boldsymbol{x})} 
\frac{\partial F}{\partial \varPi^{\varPsi}(t,\boldsymbol{x})} 
\bigg] \,.
\label{3.25}
\end{align}
In fact, this provides the Poisson brackets equivalent to Eqs. (\ref{3.22}) and (\ref{3.23}).  
Also, Eq. (\ref{3.25}) involves the Poisson brackets (\ref{3.5b}) and (\ref{3.5c}) which 
can be identified with their corresponding Dirac brackets.  
As expected, the reduced Hamiltonian (\ref{3.17}) can be written in 
terms of the canonical variables $(\varPsi, \varPi^{\varPsi})$: 
\begin{align}
\tilde{H}_{2n}
&=\int  d^{2n-1} \boldsymbol{x}  \bigg[ \,
\frac{1}{2} (M^2 +m^2) (\varPi^{P})^{2} +\frac{1}{2} P^{2} +\frac{1}{2} (\varPi^{\eta})^{2} 
+\frac{1}{2} \partial_{j}\eta \partial_{j}\eta +M\varPi^{P} \varPi^{\eta}
\nonumber 
\\
& \quad\, +\varPi^{P} (m^2 p-B) 
+\frac{1}{2m^2} \partial_{j} \varPi^{p} \partial_{j} \varPi^{p} 
+\frac{1}{2} m^2 p^{2}  
+\partial_{j} \varPi^{p} \partial_{j} \varPi^{B} 
\nonumber 
\\
& \quad\, 
+\partial_{j} \varPi^{B} \partial_{j} ( P -M \eta) 
-i \partial_{j} \varPi^{\bar{C}} \partial_{j} \varPi^{C} 
+i \bar{C} C \bigg] \,. 
\label{3.26}
\end{align}
Then, the canonical equation (\ref{3.18}) reads 
\begin{align}
\dot{F} = \{F, \tilde{H}_{2n} \}^{\ast} ,
\label{3.27}
\end{align}
where $F$ is understood as a function of $(\varPsi, \varPi^{\varPsi})$. 
Here, the weak equality symbol in Eq. (\ref{3.18}) has been replaced by the usual one, 
because the right-hand side of Eq. (\ref{3.27}) is a Poisson bracket valid on 
the phase space submanifold $\mathcal{S}$ 
and no constraints are involved in Eq. (\ref{3.27}).

The canonical equations for the canonical coordinates $(\varPsi)$ are found from 
Eq. (\ref{3.27}) to be 
\begin{subequations}
\label{3.28}
\begin{align}
\dot{P}&=(M^2 +m^2) \varPi^{P} +M\varPi^{\eta} +m^2 p -B \,, 
\label{3.28a}
\\
\dot{\eta}&=\varPi^{\eta} +M\varPi^{P} \,, 
\label{3.28b}
\\
\dot{p}&=-\Delta ( m^{-2} \varPi^{p} +\varPi^{B}) \,, 
\label{3.28c}
\\
\dot{B}&=-\Delta ( \varPi^{p} +P-M\eta ) \,, 
\label{3.28d}
\\
\dot{C}&=i \Delta \varPi^{\bar{C}} \,,
\label{3.28e}
\\
\dot{\bar{C}}&=-i \Delta \varPi^{C} \,,
\label{3.28f}
\end{align}
\end{subequations}
where $\Delta \equiv \partial_i \partial_i$. 
Similarly, the canonical equations for the momenta $(\varPi^{\varPsi})$  
are found to be  
\begin{subequations}
\label{3.29}
\begin{align}
\dot{\varPi}{}^{P} &=-P+\Delta \varPi^{B} ,
\label{3.29a}
\\
\dot{\varPi}{}^{\eta} &=\Delta( \eta-M\varPi^{B}) \,, 
\label{3.29b}
\\
\dot{\varPi}{}^{p} &=-m^2 (\varPi^{P} +p) \,, 
\label{3.29c}
\\
\dot{\varPi}{}^{B} &=\varPi^{P} ,
\label{3.29d}
\\
\dot{\varPi}{}^{C} &=i\bar{C} \,, 
\label{3.29e}
\\
\dot{\varPi}{}^{\bar{C}} &=-iC \,. 
\label{3.29f}
\end{align}
\end{subequations}
Combining Eqs (\ref{3.28}) and (\ref{3.29}) yields the equations  
\begin{subequations}
\label{3.30}
\begin{align}
&(\sq +M^{2}+m^{2}) P =0 \,, 
\label{3.30a}
\\
&\sq \eta +M P =0 \,. 
\label{3.30b}
\\
&\sq p=0 \,,
\label{3.30c}
\\
&\sq B=0 \,,
\label{3.30d}
\\
&\sq C=0 \,,
\label{3.30e}
\\
&\sq \bar{C}=0 \,. 
\label{3.30f}
\end{align}
\end{subequations}
Here, we have used  
$\sq\equiv\partial_{\mu} \partial^{\mu} =\partial^2/\partial t^2 -\Delta$. 
Equations (\ref{3.30a}) and (\ref{3.30b}) are identical to Eqs. (\ref{2.13}) and (\ref{2.12}), 
respectively. Equations (\ref{3.30c})-(\ref{3.30f}) are consistent with  
the $\nu=0$ components of Eqs. (\ref{2.16a})-(\ref{2.16d}), respectively. 
These facts imply that we have given a correct treatment of  
the present Hamiltonian system. 
The consistency of our procedure can also be seen in the BRST 
transformation rules below. 
The canonical formalism studied in this section is applied in the next section 
to quantize the fields $(\varPsi)$.

From the Lagrangian (\ref{2.8}), we can derive the BRST current,   
a Noether current associated with the BRST transformation $\boldsymbol{\delta}$. 
The BRST charge, $Q_{\rm B}$, is defined as the volume integral of 
the time component of the BRST current 
and can be written in terms of some of  the canonical variables 
$(\varPsi, \varPi^{\varPsi})$: 
\begin{align}
Q_{\rm B} =
\int  d^{2n-1} \boldsymbol{x}  \left[ 
C(\varPi^{p}+P-M\eta) +iB\varPi^{\bar{C}} \right] . 
\label{3.31}
\end{align}
Using Eqs. (\ref{3.28}) and (\ref{3.29}), we can readily show the conservation law 
$\dot{Q}_{\rm B}=0$. 
The BRST charge $Q_{\rm B}$ 
generates the BRST transformation in the following manner:  
\begin{subequations}
\label{3.32}
\begin{align}
\boldsymbol{\delta} P &= -\{ Q_{\rm B}, P \}^{\ast} =0 \,, 
\label{3.32a}
\\
\boldsymbol{\delta} \eta &= -\{ Q_{\rm B}, \eta \}^{\ast} =0 \,, 
\label{3.32b}
\\
\boldsymbol{\delta} p &= -\{ Q_{\rm B}, p \}^{\ast} =C \,, 
\label{3.32c}
\\
\boldsymbol{\delta} B &= -\{ Q_{\rm B}, B \}^{\ast} =0 \,, 
\label{3.32d}
\\
\boldsymbol{\delta} C &= -\{ Q_{\rm B}, C \}^{\ast} =0 \,, 
\label{3.32e}
\\
\boldsymbol{\delta} \bar{C} &= -\{ Q_{\rm B}, \eta \}^{\ast} =iB \,. 
\label{3.32f}
\end{align}
\end{subequations}
These are consistent with the transformation rules (\ref{2.5}) and (\ref{2.6}). 
In this way, the BRST symmetry is maintained in the reduced Hamiltonian system 
expressed in terms of the canonical variables $(\varPsi, \varPi^{\varPsi})$. 
With the aid of Eqs. (\ref{3.28d}) and (\ref{3.28e}), $Q_{\rm B}$ can be written 
\begin{align}
Q_{\rm B}=\int  d^{2n-1} \boldsymbol{x} 
\left( -C \frac{1}{\Delta} \dot{B} +B \frac{1}{\Delta} \dot{C} \right) .
\label{3.33}
\end{align}
This expression is utilized in the next section.

\section{\label{sec:level1}Canonical quantization} 

In this section, we study quantum-mechanical properties of the reduced model 
characterized by the Hamiltonian (\ref{3.26}).  
The study proceeds on the basis of the  canonical formalism developed in the previous section. 
In accordance with Dirac's quantization rule, we introduce the operators 
$F_{\rm op}$ and $G_{\rm op}$ corresponding to the functions $F$ and $G$, respectively, 
and set the (anti-)commutation relation  
\begin{align}
[ F_{\rm op}, G_{\rm op} ]_{\mp} 
&\equiv F_{\rm op} G_{\rm op} -(-1)^{|F||G|} G_{\rm op} F_{\rm op} 
\nonumber 
\\ 
&=i \{F, G\}^{\ast}_{\rm op} . 
\label{4.1}
\end{align}
Here, $\{F, G\}^{\ast}_{\rm op}$ is the operator corresponding to 
the modified Poisson bracket $\{F, G\}^{\ast}$.  
The subscript ^^ ^^ $\mp$" takes ^^ ^^ $-$" if $|F||G|=0$, 
and ^^ ^^ $+$" if $|F||G|=1$. 
The quantum-mechanical analogue of the canonical equation (\ref{3.27}) is  
the Heisenberg equation 
\begin{align}
\dot{F}_{\rm op} =-i [ F_{\rm op}, \tilde{H}_{2n \rm{op}} ]_{-} \,.
\label{4.2}
\end{align}
Hereafter, the subscript ^^ ^^ op'' is omitted for conciseness  unless confusion occurs.

From Eqs. (\ref{3.25}) and (\ref{4.1}), we have the canonical (anti-)commutation relations: 
\begin{subequations}
\label{4.3}
\begin{align}
&[ P(t,\boldsymbol{x}), \varPi^{P}(t,\boldsymbol{y}) ]_{-}
=i \delta(\boldsymbol{x}-\boldsymbol{y}) \,, 
\label{4.3a}
\\
&[ \eta(t,\boldsymbol{x}), \varPi^{\eta}(t,\boldsymbol{y}) ]_{-}
=i \delta(\boldsymbol{x}-\boldsymbol{y}) \,, 
\label{4.3b}
\\
&[\, p (t,\boldsymbol{x}), \varPi^{p} (t,\boldsymbol{y}) ]_{-}
=i \delta(\boldsymbol{x}-\boldsymbol{y}) \,, 
\label{4.3c}
\\
&[ B (t,\boldsymbol{x}), \varPi^{B} (t,\boldsymbol{y}) ]_{-} 
=i \delta(\boldsymbol{x}-\boldsymbol{y}) \,, 
\label{4.3d}
\\
&[ C (t,\boldsymbol{x}), \varPi^{C} (t,\boldsymbol{y}) ]_{+} 
=-i \delta(\boldsymbol{x}-\boldsymbol{y}) \,, 
\label{4.3e}
\\
&[ \bar{C} (t,\boldsymbol{x}), \varPi^{\bar{C}} (t,\boldsymbol{y}) ]_{+} 
=-i \delta(\boldsymbol{x}-\boldsymbol{y}) \,. 
\label{4.3f} 
\end{align}
\end{subequations}
The other canonical (anti-)commutation relations vanish. 
Using Eqs. (\ref{3.28}) and (\ref{3.29}), which are now understood as the Heisenberg 
equations, and the relations (\ref{4.3}), we can calculate the equal-time (anti-)commutation relations 
between the canonical coordinates and their time derivatives. 
Among them, all the nonvanishing relations are enumerated as follows: 
\begin{subequations}
\label{4.4}
\begin{align}
&[ P(t,\boldsymbol{x}), \dot{P}(t,\boldsymbol{y}) ]_{-}
=i (M^2 +m^2) \delta(\boldsymbol{x}-\boldsymbol{y}) \,, 
\label{4.4a}
\\
&[ P(t,\boldsymbol{x}), \dot{\eta}(t,\boldsymbol{y}) ]_{-}
=i M \delta(\boldsymbol{x}-\boldsymbol{y}) \,, 
\label{4.4b}
\\
&[ \eta(t,\boldsymbol{x}), \dot{P}(t,\boldsymbol{y}) ]_{-}
=i M \delta(\boldsymbol{x}-\boldsymbol{y}) \,, 
\label{4.4c}
\\
&[ \eta(t,\boldsymbol{x}), \dot{\eta}(t,\boldsymbol{y}) ]_{-}
=i \delta(\boldsymbol{x}-\boldsymbol{y}) \,, 
\label{4.4d}
\\
&[\, p(t,\boldsymbol{x}), \dot{p}(t,\boldsymbol{y}) ]_{-}
=-\frac{i}{m^2} \Delta \delta(\boldsymbol{x}-\boldsymbol{y}) \,, 
\label{4.4e}
\\
&[\, p(t,\boldsymbol{x}), \dot{B}(t,\boldsymbol{y}) ]_{-}
=-i \Delta \delta(\boldsymbol{x}-\boldsymbol{y}) \,, 
\label{4.4f}
\\
&[ B(t,\boldsymbol{x}), \dot{p}(t,\boldsymbol{y}) ]_{-}
=-i \Delta \delta(\boldsymbol{x}-\boldsymbol{y}) \,, 
\label{4.4g}
\\
&[ C(t,\boldsymbol{x}), \dot{\bar{C}}(t,\boldsymbol{y}) ]_{+}
=-\Delta \delta(\boldsymbol{x}-\boldsymbol{y}) \,, 
\label{4.4h}
\\
&[ \bar{C}(t,\boldsymbol{x}), \dot{C}(t,\boldsymbol{y}) ]_{+}
=\Delta \delta(\boldsymbol{x}-\boldsymbol{y}) \,. 
\label{4.4i}
\end{align}
\end{subequations}
All the equal-time (anti-)commutation relations 
between the time derivatives of the canonical coordinates vanish.

To find out physical degrees of freedom in the model, 
we need to investigate particle contents of the model. 
Before starting the investigation, 
we define a pseudoscaler field $\varphi$ by 
\begin{align}
\varphi \equiv \eta -\frac{M}{M^2 +m^2} P \,.
\label{4.5}
\end{align}
Then, from Eqs. (\ref{3.30a}) and (\ref{3.30b}), it follows that  
\begin{align}
\sq \varphi=0 \,. 
\label{4.6}
\end{align}
The transformation rules (\ref{3.32a}) and (\ref{3.32b}) guarantee 
\begin{align}
\boldsymbol{\delta} \varphi =0 \,. 
\label{4.7}
\end{align}
Using the commutation relations (\ref{4.4a})-(\ref{4.4d}), we can readily show that    
\begin{align}
[ \varphi(t,\boldsymbol{x}), \dot{\varphi}(t,\boldsymbol{y}) ]_{-}
=\frac{i m^2}{M^2 +m^2} \delta(\boldsymbol{x}-\boldsymbol{y}) \,. 
\label{4.8}
\end{align}
All the equal-time commutation relations containing either $\varphi$ or 
$\dot{\varphi}$ vanish. 
In what follows, we consider $\varphi$ to be more fundamental than $\eta$, 
because $\varphi$ satisfies the massless Klein-Gordon equation  
and simple commutation relations.

The Klein-Gordon equations (\ref{3.30a}), (\ref{4.6}) and (\ref{3.30c})-(\ref{3.30f}) 
can be solved in terms of the plane-wave basis set 
$\{e^{i\boldsymbol{k}\cdot \boldsymbol{x}} \}$: 
\begin{subequations}
\label{4.9}
\begin{align}
P(x) &=\frac{1}{(2\pi)^{(2n-1)/2}} \int \frac{d^{2n-1} \boldsymbol{k}}{\sqrt{2k_0}} 
\left\{ P(\boldsymbol{k}) e^{-ikx} +P^{\dagger} (\boldsymbol{k}) e^{ikx} \right\} , 
\label{4.9a}
\\
\varphi(x) &=\frac{1}{(2\pi)^{(2n-1)/2}} \int \frac{d^{2n-1} \boldsymbol{k}}{\sqrt{2k_0}} 
\left\{ \varphi(\boldsymbol{k}) e^{-ikx} +\varphi^{\dagger} (\boldsymbol{k}) e^{ikx} \right\} , 
\label{4.9b}
\\
p(x) &=\frac{1}{(2\pi)^{(2n-1)/2}} \int d^{2n-1} \boldsymbol{k} \sqrt{\frac{k_0}{2}}
\left\{ p(\boldsymbol{k}) e^{-ikx} +p^{\dagger} (\boldsymbol{k}) e^{ikx} \right\} , 
\label{4.9c}
\\
B(x) &=\frac{1}{(2\pi)^{(2n-1)/2}} \int d^{2n-1} \boldsymbol{k} \sqrt{\frac{k_0}{2}}
\left\{ B(\boldsymbol{k}) e^{-ikx} +B^{\dagger} (\boldsymbol{k}) e^{ikx} \right\} , 
\label{4.9d}
\\
C(x) &=\frac{1}{(2\pi)^{(2n-1)/2}} \int d^{2n-1} \boldsymbol{k} \sqrt{\frac{k_0}{2}}
\left\{ C(\boldsymbol{k}) e^{-ikx} +C^{\dagger} (\boldsymbol{k}) e^{ikx} \right\} , 
\label{4.9e}
\\
\bar{C}(x) &=\frac{1}{(2\pi)^{(2n-1)/2}} \int d^{2n-1} \boldsymbol{k} \sqrt{\frac{k_0}{2}}
\left\{ \bar{C}(\boldsymbol{k}) e^{-ikx} 
+\bar{C}^{\dagger} (\boldsymbol{k}) e^{ikx} \right\} , 
\label{4.9f}
\end{align}
\end{subequations}
where $kx\equiv k_{0} t -\boldsymbol{k}\cdot \boldsymbol{x}$. 
Here, $k_0 =\sqrt{ \boldsymbol{k}^2 +M^2 +m^2}$ for $P$, 
and $k_0 =|\boldsymbol{k}|$ for $\varphi$, $p$, $B$, $C$, and $\bar{C}$. 
Evidently, $P$ is a field with the mass $\hat{m}\equiv \sqrt{M^2+m^2}$, 
while the remainder are massless fields. 
Using the (anti-)commutation relations (\ref{4.4}) and (\ref{4.8}), 
we can derive the (anti-)commutation relations between the undetermined coefficients contained 
in Eqs. (\ref{4.9}). Among them, all the nonvanishing relations are enumerated as follows:  
\begin{subequations}
\label{4.10}
\begin{align}
&[ P(\boldsymbol{k}), P^{\dagger}(\boldsymbol{l}) ]_{-}
=(M^2 +m^2) \delta(\boldsymbol{k}-\boldsymbol{l}) \,, 
\label{4.10a}
\\
&[ \varphi(\boldsymbol{k}), \varphi^{\dagger}(\boldsymbol{l}) ]_{-}
=\frac{m^2}{M^2 +m^2} \delta(\boldsymbol{k}-\boldsymbol{l}) \,, 
\label{4.10b}
\\
&[\, p(\boldsymbol{k}), p^{\dagger}(\boldsymbol{l}) ]_{-}
=m^{-2} \delta(\boldsymbol{k}-\boldsymbol{l}) \,, 
\label{4.10c}
\\
&[\, p(\boldsymbol{k}), B^{\dagger}(\boldsymbol{l}) ]_{-}
=\delta(\boldsymbol{k}-\boldsymbol{l}) \,, 
\label{4.10d}
\\
&[ B(\boldsymbol{k}), p^{\dagger}(\boldsymbol{l}) ]_{-}
=\delta(\boldsymbol{k}-\boldsymbol{l}) \,, 
\label{4.10e}
\\
&[ C(\boldsymbol{k}), \bar{C}^{\dagger}(\boldsymbol{l}) ]_{+}
=-i\delta(\boldsymbol{k}-\boldsymbol{l}) \,, 
\label{4.10f}
\\
&[ \bar{C}(\boldsymbol{k}), C^{\dagger}(\boldsymbol{l}) ]_{+}
=i\delta(\boldsymbol{k}-\boldsymbol{l}) \,. 
\label{4.10g}
\end{align}
\end{subequations}
Equations (\ref{4.10}) are regarded as (anti-)commutation relations 
between the creation and annihilation operators for the relevant fields. 
Now we arrange the annihilation operators $(\varPsi)=(P, \varphi, p, B, C, \bar{C})$ 
and the creation operators 
$(\varPsi^{\dagger})=(P^{\dagger}, \varphi^{\dagger}, p^{\dagger}, B^{\dagger}, 
C^{\dagger}, \bar{C}{}^{\dagger})$ in the column and the row of a matrix, respectively. 
Then the (anti-)commutation relations (\ref{4.10}), 
together with the associated vanishing relations, 
can be summarized in a matrix form:  
\begin{align}
\left(\, [ \varPsi(\boldsymbol{k}), 
\varPsi^{\dagger}(\boldsymbol{l}) ]_{\mp} \right)
=\left( 
\begin{array}{cc|cccc}
M^{2}+m^{2} & 0 &\, 0 &\; 0 &\;\;\: 0 &\;\, 0 \\
0 & \dfrac{m^{2}}{M^{2}+m^{2}} &\, 0 &\; 0 &\;\;\: 0 &\;\, 0 \\ 
\hline 
0 & 0 &\, m^{-2} &\; 1 &\;\;\: 0 &\;\, 0 \\
0 & 0 &\, 1 &\; 0 &\;\;\: 0 &\;\, 0 \\ 
0 & 0 &\, 0 &\; 0 &\;\;\: 0 &\;\, -i \\
0 & 0 &\, 0 &\; 0 &\;\;\: i &\;\, 0 
\end{array}
\right) \times
\delta(\boldsymbol{k}-\boldsymbol{l}) \,.
\label{4.11}
\end{align} 
This matrix is identified with the metric matrix of the Fock subspace 
spanned by the one-particle basis vectors 
$\big\{ \varPsi^{\dagger} (\boldsymbol{k})|0\rangle \big\}$.

Substituting Eqs. (\ref{4.9d}) and (\ref{4.9e}) into Eq. (\ref{3.33}),  
we rewrite the BRST charge $Q_{\rm B}$ 
in terms of the creation and annihilation operators: 
\begin{align}
Q_{\rm B}=-i \int d^{2n-1} \boldsymbol{k} 
\left\{ C^{\dagger}(\boldsymbol{k}) B(\boldsymbol{k}) 
-B^{\dagger} (\boldsymbol{k}) C(\boldsymbol{k}) \right\} . 
\label{4.12}
\end{align}
By this procedure, $Q_{\rm B}$ is promoted to an operator. 
With Eq. (\ref{4.12}), it is easy to verify that 
$Q_{\rm B}$ generates the BRST transformation of 
the creation and annihilation operators: 
\begin{subequations}
\label{4.13}
\begin{alignat}{2}
&[iQ_{\rm B}, P(\boldsymbol{k})]_{-} =0 \,, 
&\quad 
&[iQ_{\rm B}, P^{\dagger}(\boldsymbol{k})]_{-} =0 \,,
\label{4.13a}
\\
&[iQ_{\rm B}, \varphi (\boldsymbol{k})]_{-} =0 \,, 
&\quad
&[iQ_{\rm B}, \varphi^{\dagger} (\boldsymbol{k})]_{-} =0 \,, 
\label{4.13b}
\\
&[iQ_{\rm B}, p(\boldsymbol{k})]_{-} =C(\boldsymbol{k}) \,, 
&\quad
&[iQ_{\rm B}, p^{\dagger}(\boldsymbol{k})]_{-} 
=C^{\dagger}(\boldsymbol{k}) \,, 
\label{4.13c}
\\
&[iQ_{\rm B}, B(\boldsymbol{k})]_{-} =0 \,, 
&\quad
&[iQ_{\rm B}, B^{\dagger}(\boldsymbol{k})]_{-} =0 \,, 
\label{4.13d}
\\
&[iQ_{\rm B}, C(\boldsymbol{k})]_{+} =0 \,, 
&\quad
&[iQ_{\rm B}, C^{\dagger}(\boldsymbol{k})]_{+} =0 \,, 
\label{4.13e}
\\
&[iQ_{\rm B}, \bar{C} (\boldsymbol{k})]_{+} =iB(\boldsymbol{k}) \,, 
&\quad
&[iQ_{\rm B}, \bar{C}^{\dagger} (\boldsymbol{k})]_{+} 
=iB^{\dagger}(\boldsymbol{k}) \,. 
\label{4.13f}
\end{alignat}
\end{subequations}
These are precisely the BRST transformation rules represented  
at the quantum-theoretical level. 
As easily seen, the BRST charge $Q_{\rm B}$ satisfies the nilpotency property
\begin{align}
Q_{\rm B}^{2} = \frac{1}{2} [Q_{\rm B}, Q_{\rm B} ]_{+} =0\,, 
\label{4.14}
\end{align}
and the Hermiticity condition 
\begin{align}
Q_{\rm B}^{\dagger} =Q_{\rm B} \,.
\label{4.15}
\end{align}
The transformation rules (\ref{4.13}) show that $P$ and $\varphi$ belong to 
BRST-singlet representations of the BRST algebra 
\footnote{The BRST algebra is an algebra characterized by 
Eq. (\ref{4.14}) and $[ iQ_{\rm C}, Q_{\rm B} ]_{-}=Q_{\rm B}$. 
Here, $Q_{\rm C}$ is a conserved charge associated with the scale transformation 
$C \rightarrow e^{\theta} C$,  $\bar{C} \rightarrow e^{-\theta} \bar{C}$ \cite{KO,NO}. }, 
while each of the pairs $(p, C)$ and $(\bar{C}, B)$ belongs to a BRST-doublet 
representation of this algebra.  
Considering structure of the matrix (\ref{4.11}), we see that the two doublets 
$(p, C)$ and $(\bar{C}, B)$ constitute a BRST quartet.

Using Eqs. (\ref{4.11}), (\ref{4.13}) (\ref{4.14}), and (\ref{4.15}), 
we can prove the following theorem:  
{\em $\langle f|g \rangle=\langle f|P^{(0)}|g \rangle$ is valid for 
arbitrary state vectors $|f \rangle$ and $|g \rangle$ 
satisfying $Q_{\rm B}|f \rangle =Q_{\rm B}|g \rangle =0$} \cite{KO,NO}. 
Here, $P^{(0)}$ is the projection operator onto the Fock space 
$\mathscr{H}_{\rm phys}$ 
spanned by the BRST-singlet basis vectors 
\begin{align}
\big\{ P^{\dagger} (\boldsymbol{k}_{1}) \cdots 
P^{\dagger} (\boldsymbol{k}_{a}) 
\varphi^{\dagger} (\boldsymbol{l}_{1}) \cdots 
\varphi^{\dagger} (\boldsymbol{l}_{b}) |0 \rangle \big\}_{a, b=0, 1, \ldots} \,.
\label{4.16}
\end{align}
This theorem states that in the physical subspace $\mathscr{V}_{\rm phys}$  
specified by the subsidiary condition 
\begin{align}
Q_{\rm B}|f \rangle=0 \,, 
\label{4.17}
\end{align}
the BRST-quartet particles $p$, $C$, $\bar{C}$, and $B$ are always produced 
only in zero-norm combinations and can never be observed with finite probability. 
In this way, the quartet particles appearing in $\mathscr{V}_{\rm phys}$ 
are completely confined and the Kugo-Ojima quartet mechanism 
is verified in the present model. 
Because the basis vectors (\ref{4.16}) satisfy the condition (\ref{4.17}),   
it follows that $\mathscr{H}_{\rm phys} \subset \mathscr{V}_{\rm phys}$.  
Hence the BRST-singlet particles $P$ and $\varphi$ are recognized as   
physical particles. 
In contrast to the quartet particles, the singlet particles  
may be observed with finite probability. 
To ascertain the observable particles, we investigate  
the following three cases separately:

\subsubsection{\label{sec:level3}Case $M\neq 0$, $m=0$}

In this case, the (3,3) th entry of the matrix (\ref{4.11}) diverges,   
so that the matrix (\ref{4.11}) is not well-defined. 
This is merely an apparent difficulty, giving rise to no troubles. 
In fact, we can avoid the difficulty by making the replacement 
$(p, B, C, \bar{C}) \mapsto (m^{-1}p, mB, m^{-1}C, m\bar{C})$ 
before taking $m$ to be zero. 
It should be noted that under this replacement, 
the essential properties (\ref{4.13}), (\ref{4.14}), and (\ref{4.15}) do not change at all, while  
only the (3,3) th entry of the matrix (\ref{4.11}) changes from $m^{-2}$ to $1$. 
By virtue of the replacement, the 4-by-4 submatrix in Eq. (\ref{4.11}), 
\begin{align}
\left( 
\begin{array}{cccc}
m^{-2} &\; 1 &\;\;\: 0 &\;\, 0 \\
1 &\; 0 &\;\;\: 0 &\;\, 0 \\ 
0 &\; 0 &\;\;\: 0 &\;\, -i \\
0 &\; 0 &\;\;\: i &\;\, 0 
\end{array}
\right) ,
\nonumber
\end{align} 
becomes nonsingular, and accordingly  
the theorem stated above is valid for the present case.  
Hence, the quartet particles are confined as usual  
owing to the quartet mechanism.

The commutation relations (\ref{4.10a}) and (\ref{4.10b}) in the present case 
take the following forms:  
$[ P(\boldsymbol{k}), P^{\dagger}(\boldsymbol{l}) ]_{-}
=M^2\delta(\boldsymbol{k}-\boldsymbol{l})$,  
$[ \varphi(\boldsymbol{k}), \varphi^{\dagger}(\boldsymbol{l}) ]_{-}=0$.  
These relations imply that among the basis vectors in Eq. (\ref{4.16}), 
the vectors with $\varphi^{\dagger}$ have zero norm and 
only the basis vectors $\{ P^{\dagger} (\boldsymbol{k}_{1}) 
\cdots P^{\dagger} (\boldsymbol{k}_{a}) |0\rangle \}_{a=0,1,\ldots}$  
have positive norm. For this reason, the massless singlet particle $\varphi$, 
as well as the quartet particles, can never be observed with finite probability and 
only the singlet particle $P$ with the mass $M$ can be observed.  
In other words, it can be said with Eq. (\ref{4.5}) that the massless mode 
of $\eta$ is not observable, while the massive mode of $\eta$ is observable. 
From this, it follows that $\eta$ behaves as a pseudoscalar field with the mass $M$.

\subsubsection{\label{sec:level3}Case $M=0$, $m\neq0$}

In this case, the quartet particles are, of course, confined  
due to the quartet mechanism. 
Because the right-hand sides of Eqs. (\ref{4.10a}) and (\ref{4.10b}) are together 
positive, all the basis vectors in Eq. (\ref{4.16}) have positive norm. 
For this reason, both the massless particle $\varphi$ and the particle $P$ 
with the mass $m$ can be observed. 
As seen from Eq. (\ref{4.5}), $\eta$ in this case 
is identical with $\varphi$. Hence, $\eta$ behaves as a massless pseudoscalar 
field.

\subsubsection{\label{sec:level3}Case $M\neq0$, $m\neq0$}

This case is a hybrid of the above two cases in a sense.  
The quartet particles are confined due to the quartet mechanism.  
Because the right-hand sides of Eqs. (\ref{4.10a}) and (\ref{4.10b}) are 
positive as in the case $M=0$, $m\neq0$, it follows that 
both the massless particle $\varphi$ and the particle $P$ with the mass 
$\hat{m}\equiv \sqrt{M^2+m^2}$ can be observed. 
This implies that $\eta$ can behave as a massive pseudoscalar field with the mass 
$\hat{m}$ \cite{DH}.  

\vspace{7mm}

In all the three cases, 
the particle $P$ is recognized as the only massive particle 
that can be observed with finite probability. 
The massless particle $\varphi$ is recognized as an observable particle 
if and only if $m\neq0$.

\section{\label{sec:level1}Converting to the modified hybrid model} 

The BRST transformation rule of the Yang-Mills fields $A_{\mu}^{a}$ is defined 
by replacing the parameters $\omega^{a}$ in Eq. (\ref{1.3}) by the FP ghost fields 
$c^{a}$: 
\begin{align}
\boldsymbol{\delta} A_{\mu}^{a} =D_{\mu} c^{a} . 
\label{5.1}
\end{align}
Here, $c^a$ are, of course, anticommutative fields. 
The nilpotency property $\boldsymbol{\delta}^{2}=0$ is maintained by setting 
the transformation rule 
$\boldsymbol{\delta} c^{a}=\frac{1}{2}g f_{bc}{}^{a} c^{b} c^{c}$. 
The BRST transformation rule of the Chern-Simons current 
$\mathcal{C}_{2n}^{\mu}$ is found from Eq. (\ref{1.4}) to be 
\begin{align}
\boldsymbol{\delta} \mathcal{C}_{2n}^{\nu} 
=\partial_{\mu} \mathcal{C}_{2n}^{\mu\nu} \,, 
\label{5.2}
\end{align}
where $\mathcal{C}_{2n}^{\mu\nu}$ is defined by replacing $\omega^{a}$ 
included in $\mathcal{U}_{2n}^{\mu\nu}$ by $c^{a}$: 
$\mathcal{C}_{2n}^{\mu\nu}\equiv \mathcal{U}_{2n}^{\mu\nu}|_{\omega^{a}=c^{a}}$. 
The BRST transformation rule of $\mathcal{C}_{2n}^{\mu\nu}$ is 
determined to be 
\begin{align}
\boldsymbol{\delta} \mathcal{C}_{2n}^{\mu\nu} 
=\partial_{\rho} \mathcal{C}_{2n}^{\rho\mu\nu} , 
\label{5.3}
\end{align}
where $\mathcal{C}_{2n}^{\rho\mu\nu}$ is a rank-3 totally antisymmetric tensor 
that is a polynomial in $(A_{\mu}^{a}, F_{\mu\nu}^{a}, c^{a})$ and 
quadratic in $c^{a}$. 
Using Eq. (\ref{5.3}) and the antisymmetry property of 
$\mathcal{C}_{2n}^{\rho\mu\nu}$ in its indices, it can be shown    
that $\boldsymbol{\delta}^{2} \mathcal{C}_{2n}^{\nu}=0$. 
Equations (\ref{5.2}) and (\ref{5.3}) are precisely constituents of 
the chain of descent equations 
$\boldsymbol{\delta} \mathcal{C}_{2n}^{\nu_{1}\cdots \nu_{p}} 
=\partial_{\mu} \mathcal{C}_{2n}^{\mu\nu_{1}\cdots \nu_{p}}$ 
$(p=1,2,\ldots,2n)$, with $\mathcal{C}_{2n}^{\mu\nu_{1}\cdots \nu_{2n}}=0$ \cite{CS}. 
Here, $\mathcal{C}_{2n}^{\nu_{1}\cdots \nu_{p}}$ is a rank-$p$  
totally antisymmetric tensor. 
The relation $\boldsymbol{\delta}^2 \mathcal{C}_{2n}^{\nu_{1}\cdots \nu_{p}}=0$ is valid 
by virtue of antisymmetry of $\mathcal{C}_{2n}^{\nu_{1}\cdots \nu_{p+2}}$ in its indices. 
In this way, the nilpotency of $\boldsymbol{\delta}$ is guaranteed with 
the chain of descent equations.

Now, let us introduce an anticommutative vector field $\bar{\varGamma}_{\mu}$ and 
a commutative vector field $\mathcal{B}_{\mu}$ that obey  
the BRST transformation rules 
\begin{subequations}
\label{5.4}
\begin{align}
\boldsymbol{\delta} \bar{\varGamma}_{\mu} &=i \mathcal{B}_{\mu} , 
\label{5.4a}
\\
\boldsymbol{\delta} \mathcal{B}_{\mu} &=0\,. 
\label{5.4b}
\end{align}
\end{subequations}
Obviously, these satisfy the nilpotency property $\boldsymbol{\delta}^{2}=0$. 
We consider a BRST-coboundary term 
\begin{align}
\mathcal{L}_{K\mathcal{C}}=i\boldsymbol{\delta}\big[ 
\bar{\varGamma}_{\mu} ( K^{\mu} -\mathcal{C}_{2n}^{\mu}) \big] \,, 
\label{5.5}
\end{align}
which can be written, after the use of Eqs. (\ref{2.5a}), (\ref{5.2}) and (\ref{5.4a}), as 
\begin{align}
\mathcal{L}_{K\mathcal{C}}=-\mathcal{B}_{\mu} ( K^{\mu}-\mathcal{C}_{2n}^{\mu})
-i\bar{\varGamma}_{\nu} \partial_{\mu} 
(C^{\mu\nu}-\mathcal{C}_{2n}^{\mu\nu}) \,.
\label{5.6}
\end{align}
Adding $\mathcal{L}_{K\mathcal{C}}$ to Eq. (\ref{2.8}), we have the new Lagrangian 
\begin{align}
\tilde{\mathcal{L}}_{2n}& \equiv  \hat{\mathcal{L}}_{2n} +\mathcal{L}_{K\mathcal{C}} 
\nonumber 
\\
& = -\frac{1}{2} P^2 -\frac{1}{2} m^{2} (K^{\nu} -\partial_{\mu} p^{\mu\nu}) 
(K_{\nu} -\partial^{\rho} p_{\rho\nu}) 
\nonumber 
\\ 
& \quad \,
-K^{\mu} \partial_{\mu} (P -M\eta )
+\frac{1}{2} \partial_{\mu} \eta \partial^{\mu} \eta 
\nonumber 
\\ 
& \quad \, 
-K^{\mu} \partial^{\nu} B_{\nu\mu} -{\alpha\over 4} B_{\mu\nu} B^{\mu\nu}
-i \partial_{\mu} \bar{C}^{\mu\nu} \partial^{\rho} C_{\rho\nu} 
\nonumber \\ 
&\quad \, -\mathcal{B}_{\mu} ( K^{\mu}-\mathcal{C}_{2n}^{\mu})
-i\bar{\varGamma}_{\nu} \partial_{\mu}
 (C^{\mu\nu}-\mathcal{C}_{2n}^{\mu\nu}) \,.  
\label{5.7} 
\end{align}
From $\tilde{\mathcal{L}}_{2n}$, 
the Euler-Lagrange equations for $P$, $\mathcal{B}_{\mu}$ and $\bar{\varGamma}_{\nu}$ 
are found to be  
\begin{subequations}
\label{5.8}
\begin{align}
&P=\partial_{\mu} K^{\mu}, 
\label{5.8a}
\\
&K^{\mu}=\mathcal{C}_{2n}^{\mu}, 
\label{5.8b}
\\
&\partial_{\mu}C^{\mu\nu}=\partial_{\mu} \mathcal{C}_{2n}^{\mu\nu}. 
\label{5.8c}
\end{align}
\end{subequations}
Combining Eqs. (\ref{5.8a}) and (\ref{5.8b}) leads to  
$P=\partial_{\mu} \mathcal{C}_{2n}^{\mu} =\mathcal{P}_{2n}$, and therefore    
the field $P$ can be identified with the Chern-Pontryagin density $\mathcal{P}_{2n}$. 
Using Eqs. (\ref{5.8}), the fields $P$, $K^{\mu}$, and $\partial_{\mu}C^{\mu\nu}$ 
can be eliminated from Eq. (\ref{5.7}); after the elimination, 
$\tilde{\mathcal{L}}_{2n}$ is equivalently written as 
\begin{align}
\tilde{\mathcal{L}}_{2n}^{\rm top} &= \frac{1}{2} \mathcal{P}_{2n}^{2} 
-\frac{1}{2} m^{2} (\mathcal{C}_{2n}^{\nu} -\partial_{\mu} p^{\mu\nu}) 
(\mathcal{C}_{2n,\nu} -\partial^{\rho} p_{\rho\nu}) 
\nonumber 
\\
& \quad \,
-M\eta \mathcal{P}_{2n} 
+\frac{1}{2} \partial_{\mu} \eta \partial^{\mu} \eta 
\nonumber 
\\
& \quad \,
- \mathcal{C}_{2n}^{\mu} \partial^{\nu} B_{\nu\mu} 
-{\alpha\over 4} B_{\mu\nu} B^{\mu\nu}
-i \partial^{\mu} \bar{C}_{\mu\nu} 
\partial_{\rho} \mathcal{C}_{2n}^{\rho\nu}
\label{5.9} 
\end{align}
up to a total derivative term. 
This is precisely the Lagrangian (\ref{1.7}) supplemented with 
a sum of gauge-fixing and FP ghost terms.    
Evidently, the Lagrangian $\tilde{\mathcal{L}}_{2n}^{\rm top}$ is BRST invariant. 
The equivalence between $\tilde{\mathcal{L}}_{2n}$ and $\tilde{\mathcal{L}}_{2n}^{\rm top}$ 
can also be proven at the quantum-theoretical level via the path integrations over 
$P$, $\mathcal{B}_{\mu}$, and $\bar{\varGamma}_{\nu}$ in the generating functional with 
the Lagrangian $\tilde{\mathcal{L}}_{2n}$. 
Thus, the equivalent model (with the gauge-fixing term (\ref{2.7})) 
is converted to the modified hybrid model 
(with a corresponding gauge-fixing term) by incorporating 
the BRST-coboundary term $\mathcal{L}_{K\mathcal{C}}$ into the equivalent model. 
Because the Lagrangians of the two models, 
$\hat{\mathcal{L}}_{2n}$ and $\tilde{\mathcal{L}}_{2n}$, are connected via   
a BRST-coboundary term in such a manner that 
$\tilde{\mathcal{L}}_{2n}=\hat{\mathcal{L}}_{2n}+\mathcal{L}_{K\mathcal{C}}$,  
the two models are considered to be equivalent in the BRST-cohomological sense.   
As a result, the two models are classified into the same cohomology class.

With the identification $P=\mathcal{P}_{2n}$, 
we can conclude from the fact stated in the last part of 
Sec. 5 that the Chern-Pontryagin density $\mathcal{P}_{2n}$ behaves as an observable 
pseudoscalar particle (or field) with the mass $\hat{m}$.  
This is consistent with a result of the classical analysis made in 
Ref. \onlinecite{DH}. 
The topological mass generation in the modified hybrid model 
is thus verified at the quantum-theoretical level.  
Another relevant BRST-singlet field is the massless pseudoscalar field 
\begin{align}
\varphi \equiv \eta -\frac{M}{M^2 +m^2}  \mathcal{P}_{2n}\,.
\label{5.10}
\end{align}
If $m\neq 0$, $\varphi$, as well as $\mathcal{P}_{2n}$, can be observed 
with finite probability. 
If $m=0$, $\varphi$ can never be observed and 
only $\mathcal{P}_{2n}$ can be observed. 
It should be stressed here that the possibility of observation 
of $\mathcal{P}_{2n}$ and $\varphi$ 
can be examined only in the quantum-theoretical framework; 
it cannot be discussed at the classical level. 
In Ref. \onlinecite{DJP}, Dvali {\it et al.} considered, at the classical level, 
a model with the axial vector current $\mathcal{J}_{\mu}^{5}=\partial_{\mu} \eta$. 
In terms of the current formulation, 
this model is read as the case $m=0$ in four dimensions.  
At present, it is clear that $\mathcal{P}_{4}$ is the only observable in their model.

\section{\label{sec:level1}Summary and discussion} 

We have studied the canonical formalism for the modified hybrid model,  
aiming at clarifying particle contents of the model. 
To avoid treating the constituent Yang-Mills fields, 
the canonical formalism itself was considered for a model that is equivalent to 
the modified hybrid model but does not contain the constituent Yang-Mills fields. 
The equivalence here was established owing to the fact that 
the Lagrangian of the equivalent model, Eq. (\ref{1.12}),  
has the same form as that of the modified hybrid model, Eq. (\ref{1.7}).

The equivalent model possesses an Abelian gauge symmetry with 
a pseudotensorial gauge parameter. 
To fix the gauge of this symmetry, 
the BRST gauge-fixing procedure was adopted for convenience.  
After that, the canonical formalism of the equivalent model was considered,  
in which the Dirac formulation of constrained Hamiltonian systems 
was applied to dealing with the constraints arising in the model.  
The constraints were treated as strong equations using the Dirac bracket, 
and some of them, Eq. (\ref{3.11}), were solved  
in terms of canonical momenta of the pseudoscalar type. 
The Hamiltonian system was simply described by using these momenta and 
their conjugate pseudoscalar fields.  
In fact, the reduced Hamiltonian took the simple form of Eq. (\ref{3.26}).

The canonical quantization of the equivalent model was performed on 
this Hamiltonian system in accordance with Dirac's quantization rule. 
Thereby the particle contents of the equivalent model were clarified,  
and each of the particles was classified into a BRST singlet or quartet particle.  
It was shown that 
the two BRST-singlet particles $P$ and $\varphi$, 
which are massive and massless, respectively, are present in the model 
as genuinely physical particles.  
From the commutation relations of the BRST-singlet particles, 
it was found that $P$ can be observed with finite probability, 
provided that the mass parameters $M$ and $m$ do not vanish simultaneously. 
It was also found that the massless particle $\varphi$ can be observed  
with finite probability if and only if $m\neq0$.

The equivalent model (with the gauge-fixing term (\ref{2.7})) 
was converted to the modified hybrid model (with a corresponding gauge-fixing term) 
in a BRST-invariant manner by incorporating the BRST-coboundary term (\ref{5.5}) 
into the equivalent model. 
Through this procedure, the equivalence of the two models was 
established in the BRST-cohomological sense. 
Also, the massive particle $P$ in the equivalent model was identified 
with the Chern-Pontryagin density $\mathcal{P}_{2n}$ in the modified hybrid model. 
As a result, $\mathcal{P}_{2n}$ was recognized as an observable pseudoscalar 
particle (or field) with the mass $\hat{m}$. 
In this way, the topological mass generation studied in 
Refs. \onlinecite{DJP} and \onlinecite{DH} was 
shown for the modified hybrid model at the quantum-theoretical level.

It has been stated in Ref. \onlinecite{DH} that the modified hybrid model in four dimensions 
should have a close connection with the effective Lagrangian approach \cite{effect} 
to the U(1) problem in quantum chromodynamics (QCD). 
In fact, the modified hybrid model with $m=0$,  
or rather the equivalent model with $m=0$, was considered before 
in the effective Lagrangian approach in order to phenomenologically describe 
the generation of a large $\eta^{\prime}$ mass.   
In this approach, the field equation $\partial^{\mu} \mathcal{J}_{\mu}^{5}=-MP$ 
($\mathcal{J}_{\mu}^{5} \equiv \partial_{\mu} \eta$) derived from Eq. (\ref{2.2}) 
is understood as the hadronic analogue  
to the anomalous conservation law of the axial vector current consisting of the quark fields.   
The field $P$ is then identified with the Chern-Pontryagin density $\mathcal{P}_{4}$.  
Considering $\mathcal{P}_{4}$ as a composite state of the Yang-Mills fields $A_{\mu}^{a}$ 
that represent gluons, we can interpret $P$ as the pseudoscalar glueball field \cite{PBPRV}. 
Correspondingly, the equivalent model with $m=0$ can be 
regarded as a phenomenological model that treats 
the pseudoscalar glueball as well as the $\eta^{\prime}$ meson. 
Now, recall that the identification $P=\mathcal{P}_{4}$ is involved in the 
Lagrangian $\tilde{\mathcal{L}}_{4}$, namely Eq. (\ref{5.7}) with $n=2$.  
Noting this remarkable fact, 
we can consider $\tilde{\mathcal{L}}_{4}$ to be appropriate for describing 
the $\eta^{\prime}$ mass generation in the QCD inspired model.

The modified hybrid model with $m\neq0$ 
was first proposed in Ref. \onlinecite{DH} and has not been applied to 
phenomenology yet. 
It seems that 
the existence of the observable massless particle 
$\varphi$ causes some difficulties in phenomenological applications of 
the modified hybrid model with $m\neq0$. 
Such difficulties would be overcome by introducing an extra gauge field, 
because the massless field $\varphi$ may be absorbed into the extra gauge field 
in a manner similar to the Higgs mechanism. 
Details of this possibility should be discussed in the future. 

\begin{acknowledgments}
The author would like to thank Professor. K. Fujikawa for his encouragement   
and useful comments.  
\end{acknowledgments}

\end{document}